# Enabling the Long-Term Archival of Signed Documents through Time Stamping

Petros Maniatis T.J. Giuli Mary Baker
{maniatis,giuli,mgbaker}@cs.stanford.edu
http://identiscape.stanford.edu/
Department of Computer Science
Stanford University
Stanford, California 94305-9040

1st November 2018

## Abstract

In this paper we describe how to build a trusted reliable distributed service across administrative domains in a peer-to-peer network. The application we use to motivate our work is a public key time stamping service called Prokopius. The service provides a secure, verifiable but distributable stable archive that maintains time stamped snapshots of public keys over time. This in turn allows clients to verify time stamped documents or certificates that rely on formerly trusted public keys that are no longer in service or where the signer no longer exists. We find that such a service can time stamp the snapshots of public keys in a network of 148 nodes at the granularity of a couple of days, even in the worst case where an adversary causes the maximal amount of damage allowable within our fault model.

## 1 Introduction

Recent efforts in peer-to-peer networking [CSWH00, DFM00, WRC00, Gnu, KBC$^+$00] have demonstrated the feasibility of providing very large-scale distributed services. Different peer-to-peer networks provide various combinations of many valuable traits, such as anonymity, data permanence, efficient queries, authentication, and the ability to include participants from many different organizations and countries with no single central authority. This last trait is of particular importance to applications that offer reliability guarantees. As an example, the resistance of a document to censorship or our belief in the integrity of a document's content may be more achievable for political reasons when the network supporting the document storage and retrieval is composed of independent nodes in many countries with the nodes able to come to a majority agreement about the existence or contents of the document.

Yet without a single administrative authority or source of trust information about nodes, it remains difficult in these networks to build such services when the external information required for validating query results is no longer available. This could occur, for instance, if the validity of the document resulting from a query relies on a digital signature whose signing or verification key is no longer valid or available.

In this paper, we explore how to build long-lived reliable trusted services in a very loosely-coupled, distributed environment. In particular, we look at how to enable participating servers from diverse, independent administrative domains to implement a communal service that can be considered as a "trusted third party," even if not all participating servers are trusted by every user of this service. We also describe ways to preserve the continuity of such a service across membership or network changes over time as participants join, fail, or prove themselves untrustworthy.

The application we use to motivate this effort is a distributed secure public key time-stamping service, called Prokopius. Prokopius builds a secure, verifiable but distributable stable archive that maintains time stamped snapshots of public keys over time. This in turn allows clients to verify time stamped documents or certificates that rely on formerly trusted public keys that are no longer in service. Such functionality is essential when archiving digitally signed documents whose validity is dependent on an expired signature, not only after the signing key has changed, but even after the signer is no longer available.

While very useful, such a service would have to be very thoroughly trusted. To our knowledge, only centralized implementations of time stamping services exist today [Sur]. We believe that distributing not only the functionality, but also the organizational administration of such a service, can increase its ability to address the trust requirements of the diverse global digital community that the Internet has created. Although we motivate our work with Prokopius, the work is also valid for any service that requires the long-term archival of authenticated information.

We evaluate the agreement protocols used in Prokopius using a new simulator, called Narses, that simulates transport-layer flows, eliding individual packet information. This is faster than packet-level simulators [FV99] for large-scale networks, while still capturing the interdependences of communications. Our results show that time stamping rounds on the order of a couple of days are entirely feasible in a network of 148 nodes, even where an adversary causes the maximum amount of damage possible within our fault model. Given the slow rate of change of public key information, this is appropriate for our application domain, although it may be too slow for other applications. While slower than a centralized service, the distributed service is more believable and more survivable, across both natural disasters and .com failures.

The overall contributions of this paper are

- The design of a secure distributed public key time stamping service that is survivable even across a complete change in service provider membership over time,
- A deployable implementation of protocols for Byzantine agreement across different administrative domains,
- The definition of ways in which these protocols can be used in the face of node membership changes and only locally managed trust relationships between nodes,
- A simulator based on transport-layer flows eliding individual packet information, and
- A performance evaluation of these protocols in the context of Prokopius.

The rest of this paper is organized as follows: In Section 2 we describe the current state of the art for time stamping services and for long-term archival of signed documents. In Section 3 we go over the functionality of Prokopius in the larger context of time stamping services on the Internet through a detailed walk-through. In Section 4 we describe in more detail the actual tasks performed by Prokopius, after listing explicitly the assumptions we make. Section 5 presents the component protocols we use as building blocks. In Section 6 we describe the Narses flow-based simulator. In Section 7 we evaluate Prokopius based on the time required to execute its core components, given the maximum damage an adversary may cause. In Sections 8 and 9 we describe related and future work, respectively. We conclude in Section 10.

# 2 State of the art

In this section we outline the relevant facts of time stamping and digital signatures, on which the Prokopius system is based. First we begin with a description of how centralized time stamping works. Then we briefly list the objectives of digital signatures, the implications of their limited lifespan, and how time stamping can fortify them for long term storage.

## 2.1 Time stamping

Digital time stamping is a service whose objective is to fortify documents in digital form with a cryptographic guarantee of their minimum and maximum age, as well as their integrity [HS91].

A *Time Stamping Service* (TSS) accepts documents from clients for time stamping. In response to a submitted document, the TSS returns to the requester a *time stamp*, which is a cryptographic value derived from the document. The derivation of the value uses, among other things, a collision-resistant, one-way function; this makes it intractable to find a different, meaningful document that matches the same time stamp as the original document. Consequently, given a valid time stamp, the integrity of a document can be protected.

The time stamp also gives anyone who cares to verify it a guarantee on the age of the document. The actual guarantee varies from service to service, and from time stamping scheme to time stamping scheme. In its simplest form, the time stamp contains a time designation (i.e., date and time), and guarantees that "the document existed in this form at the time included in the time stamp". In the most naïve implementation of this guarantee, the TSS merely signs the above quoted statement for every document it time stamps; as long as everybody trusts the TSS, such signed statements are sufficient.

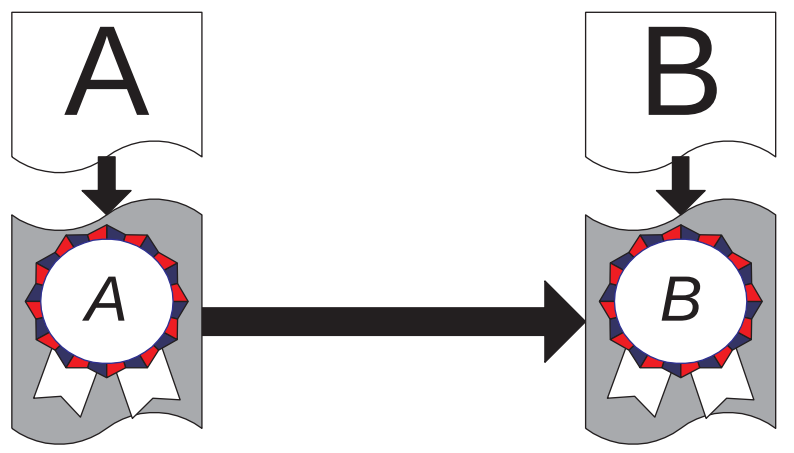

Figure 1: The time stamp for document B is dependent on the contents of document B and on the time stamp for document A.

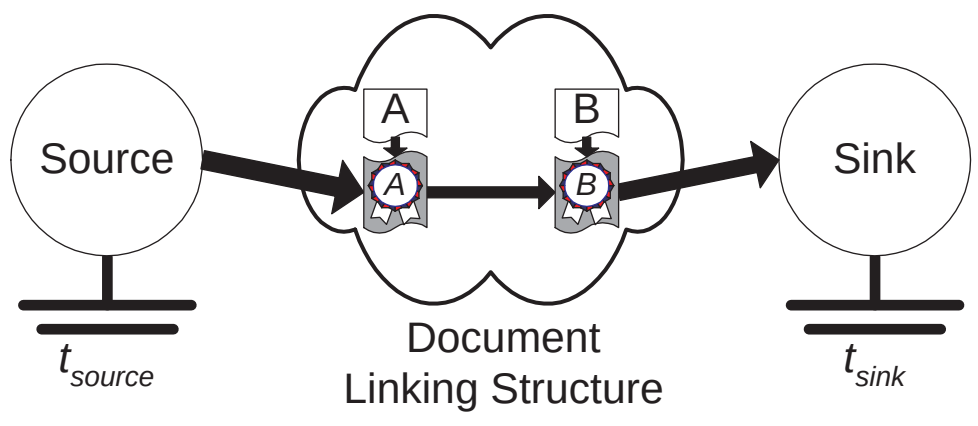

Figure 2: A binary linked structure grounded on well-known times $t_{source}$ and $t_{sink}$. Since every document in the structure could only be time stamped after the value in the *Source* was determined, and the value in the *Sink* could only be determined after all documents in the structure were time stamped, the times of the *Source* and *Sink* can place the time stamping of all documents in the structure within the time interval $[t_{source}, t_{sink}]$.

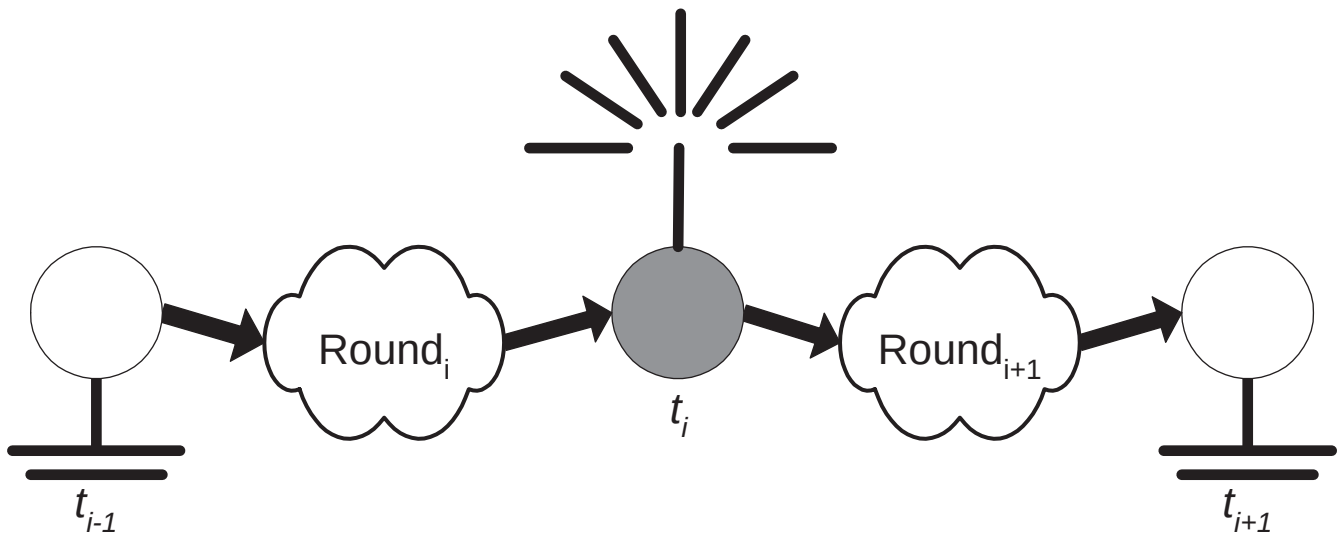

Figure 3: Two consecutive rounds. Round $i$ contains documents time stamped between times $t_{i-1}$ and $t_i$. Round $i+1$ contains documents time stamped between times $t_i$ and $t_{i+1}$. All documents in round $i$ predate (as far as time stamping goes) all documents in round $i+1$. The shaded value is not grounded in a widely witnessed way, whereas the values for time points $t_{i-1}$ and $t_{i+1}$ are.

However, most practical TSSes also offer their clients guarantees of accountability. The TSS takes extra steps to convince its clients that it cannot "cheat", i.e., it cannot back- or post-date documents surreptitiously. This is accomplished using *binary linking schemes* [BdM91, BLLV98]. These schemes link documents together cryptographically in directed graph structures so that arcs in the graph follow the time ordering of the stamping operations (i.e., if there is an arc from document A to document B, then document A was time stamped before document B). Linking schemes rely on collision-resistant, one-way functions as well. In Figure 1, the time-stamp for document B is derived from document A using such a function, and the linking structure is a singly linked list. From the derivation, we can safely assume that the TSS had possession of document A when it time stamped document B; that is, document A existed before document B was time stamped by the TSS. If, instead, the TSS did not know A or its time stamp when it time stamped B, then it must have been able to find a document A that maps – along with document B – to the time stamp of B after the fact, which is intractable given the assumptions about the computation of time stamps and one-way, collision-resistant functions.

Binary linking structures create precedence relationships among documents — from document A to document B, in the example — but say nothing about the absolute times at which those documents existed. To place the creation of the time stamp of document B at a well-known point in time, one could publish it in a popular newspaper. Then it is simple to argue that document A existed before the time of publication of the newspaper issue, which places the document in time. TSSes do exactly that. First they link documents submitted roughly at the same time in a connected linking structure with a single source and a single sink vertex (i.e., a single vertex with no incoming arcs and a single vertex with no outgoing arcs). Then they "ground" the source and sink vertices to well-known points in time through publication, thereby placing in time all intervening vertex-documents (see Figure 2).

Documents participating in the same linking structure make up a *time stamping round* (see Figure 3). Rounds are linked to each other through their sources and sinks. In the figure, round $i+1$ uses as its source the sink of its immediately preceding round $i$. Therefore, all documents in rounds $i$ and earlier transitively precede documents in rounds $i+1$ and later.

In practice, there is a gap between the frequency with which "grounding" occurs and that with which rounds change. This is because the publication of grounding values is fairly expensive and time consuming – it involves taking out an ad in a popular newspaper, or widely dis-tributing a write-once medium like a CD-ROM. On the other hand, since round duration determines the granularity of time stamping, for most practical applications it has to be kept relatively short. Typically, time stamping rounds come at the granularity of a second, but grounding occurs only at the granularity of a week [Sur]. Intra-publication source/sink vertex values, i.e., the values that do not get the wide, write-once distribution of the infrequent grounding values, are broadcast by the TSS on-line and are taken by the clients "on faith". This is not a major problem, since grounding protects only against the "unlikely" event that the TSS cheats. This frequency disparity makes it possible for the TSS to cheat, but only between grounding value publications, which is a period of a week or so. A malicious TSS could post- or pre-date a document by a few hours, even a couple of days, but not by more than a week. In the example of Figure 3, where the grounding value between rounds $i$ and $i+1$ is *not* published indisputably, a cheating TSS could pre-date a document time stamped at time $t_i \leq t \leq t_{i+1}$ to look as if it were stamped at time $t_{i-1} \leq t' \leq t_i$. However, it still cannot make the document look as if it were stamped after $t_{i+1}$ or before $t_{i-1}$, since it cannot change the published grounding values before round $i$ and after round $i+1$.

In the explanation above, we have described the time stamp of a document vaguely, as a cryptographic value derived from the preceding documents. In fact, the time stamp also contains information that can be used to prove that the document fits within its linking structure. Using this information, a verifier can ascertain that the unforgeable link from the grounded source to the document to the grounded sink holds. The party who submitted the document to the TSS for time stamping can also check the validity of the linkage right after the associated sink is made public, to ensure that the TSS performed its task correctly. In the example described in Figures 1 and 2, the "proof" of document A's time stamp consists of document B, given source and sink values whose authenticity is undisputed. A verifier can derive A's time stamp from the source and A's contents, then B's time stamp from A's time stamp and B's contents, and then the sink from B's time stamp. If the time stamp for B that the verifier derives matches the one he was given, and if the derived sink value matches the public sink value, the verifier accepts the time stamp.

In practice, the "proof" portion of a time stamp does not really have to contain the entire documents related to the time stamped document; in fact, it should not contain entire documents, for privacy and efficiency reasons. A proof need only contain a *fingerprint* of those related documents. A fingerprint of a document is a cryptographic digest of that document, a value derived from it using a one-way, collision-resistant function, which represents the document securely.

Several efficient binary linking schemes have been proposed. In the scheme employed in this paper, document digests and a Merkle tree [Mer80] structure are used, as first proposed within a time stamping setting by Benaloh in [BdM91]. The resulting time stamps have size proportional to $\log n$, where $n$ is the number of documents in a single linking structure. Details are presented in Section 5.6.

To perform the validity checks on document integrity, time placement and the time stamp described above, a verifier must have available the document itself, the time stamp on that document, and the grounding values (sink and source) immediately preceding and following the round of that document.

Although in most current time stamping systems, it is assumed that a verifier can obtain the grounding values necessary for a time stamp verification, it is not clear how that can be done in the general distributed case, where multiple TSSes coexist, and no single one of them is trusted by everyone. It is essential for a verifier to validate the grounding values themselves, to verify, that is, that they are unique for the given TSS and the chosen time periods. Otherwise, a malicious TSS or publisher could supply the verifier with any grounding value that would allow a faulty time stamp to be accepted. Locating appropriate grounding values is a time stamping problem by itself, and can be fairly complex when time stamps must persist for long periods of time (several decades) and must convince verifiers in different localities from the original document submitter. The survivability of the secure, write-once medium used to publish grounding values is of paramount importance for the longevity of a time stamping system.

One of the goals of the work presented in this paper is to address time stamp *survivability*. We extend the basic time stamping model to allow time stamps to maintain their validity even after the TSS that issued them goes out of business. Our extended time stamping model is based on a globally-trusted time stamping infrastructure, that does not require global trust of any one of its basic components, and consequently, does not require a globally-trusted, immortal publisher of grounding values for its wide-spread acceptance. We know of no other time stamping service that solves these two problems.

## 2.2 The lifetime of digital signatures

In many ways, time stamps and digital signatures provide similar services. Time stamps authenticate a document *temporally*, by certifying the time frame during which a document was dated. Similarly, digital signatures authenticate an endorsement on a document, by certifying the identity of the party who endorsed that document.

Briefly, if entity $A$ wants to sign document $D$, $A$ performs an unforgeable but verifiable computation $s_k(.)$ on the digest $d = digest(D)$ of $D$, using key $k$. This computation produces the signature $\sigma = s_{S_A}(digest(D))$; $S_A$ is the secret signing key of $A$ and it is intractable for anyone who does not have that key to create such signatures anew on any documents. Anyone in possession of $A$'s public key $P_A$ can verify that a signed document was in fact signed by someone who knew $S_A$.

Unfortunately, cryptographic keys have limited lifetimes. Whether to limit the potential damage in case of secret key leakage, or to thwart the efforts of cryptanalytic attacks, a signing key should be proactively discarded and replaced by a fresh one either after it has been used to sign a large number of documents, or after it has been in use for a long period of time. Exact thresholds on the signing volume or usage period of a signing key vary, depending on the level of security required (see [LV99] for an analytical approach to this problem). Furthermore, when a key compromise is explicitly detected, a new key should be created immediately.

The ephemeral nature of signing keys makes the task of archiving signed documents complex. Although $A$'s document $D$ might well be properly signed using $S_A$ when $A$ first creates it, it is unclear how a verifier can check the validity of the signature after $A$ has changed his signing key to ${S_A}'$. For instance, in the case that $D$ represents a strenuous contractual obligation undertaken by $A$ (to which $A$ testifies, by signing $D$), if $A$ can claim at any time that his key has been compromised, then he can effectively repudiate his contractual obligations outlined in $D$. The converse problem arises when $A$'s compromised key is used to sign an unwanted statement $D'$. How can $A$ disclaim his apparent signing of $D'$ if there is no telling whether the document was signed before or after the compromise of $S_A$?

For these reasons, time stamping can be essential. A signed and then time stamped document that predates a change of signing keys by the signer is arguably non-repudiable. Although a thief of $A$'s key can sign arbitrary documents, she cannot have them time stamped so as to predate a time stamped report of key change, as long as the time stamping service maintains its integrity.

It should be noted that if a key is compromized and used *before* it has been reported changed, then its illicit time stamped use is indistinguishable from its regular use by its rightful owner. This is an inherent shortcoming of the currently available digital signing facilities. It can only be combatted through strict adherence to a proactive key change regime, which can curtail the amount of damage in which a potential key compromize can result, and through the use of different keys for signing and for secure communication with the public key infrastructure.

Though time stamping can strengthen signed documents to defy the inherently limited life of signing keys, it does not suffice by itself. A verifier seeking to check the time stamped signature on document $D$ would have to know which particular public signing key to use. This problem does not have a trivial solution. For example, bundling the appropriate current public key with the signed document before time stamping would not work. To prove that the bundled public key is the right one (that is, the key assigned to the signer by the responsible Certification Authority), the signer would have to include an *identity certificate* $C_A$, a statement equivalent to "Key $P_A$ belongs to entity $A$", signed by the Certification Authority. Then, the public key of the Certification Authority would also have to be included in the package. Unfortunately, since Certification Authorities also come and go, and invariably identity certificates for such authorities are self-signed, no certificate for the Certification Authority is necessarily resistant to aging. In essence, we would like a "grounding" identity certificate stating "Key $P_{CA}$ belongs to entity $CA$ at time $t_0$" that we could trust forever, to bootstrap any signature verification process.

In this paper, we propose a key archival service that produces exactly such boot-strapping identity certificates. The service also provides the properties of trust continuity and potentially global trust, and it runs in conjunction with a distributed time stamping service. The following section presents a functional description of Prokopius.

# 3 Prokopius in perspective

In this section we introduce Prokopius, a basic infrastructure service that enables solutions to the longevity and global acceptance problems faced by archives of signed documents as described above. We examine Prokopius from the functional standpoint, first by giving an overview of the functionality it provides, and then by walking through an example of its use and role within a larger archival service.

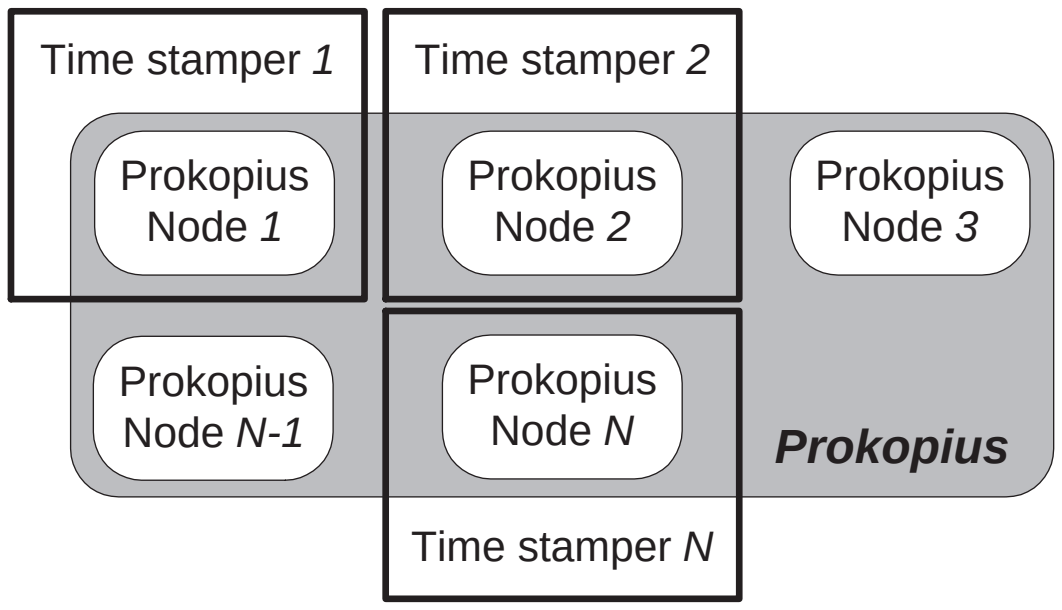

Figure 4: Relationship among Time Stamping Services and Prokopius. Every time stamping service participates in Prokopius through its own Prokopius node. Some Prokopius nodes might not belong to a Time Stamping Service. The shaded region depicts the Prokopius system.

## 3.1 Functional interface

The objective of the Prokopius system is to make wide-area, long-term, secure archives of signed documents possible through the use of time stamping and identity certificate archival.

Prokopius runs on a large set of nodes (we envision systems of the size of a few hundred nodes). All or most of the participating nodes are operated by distinct TSSes. Those services run almost exactly as current, traditional TSSes do, independently of other Prokopius nodes. However, instead of publishing their "grounding values" to a newspaper or other timed write-once medium, these time stamping services use as their publication medium the peer-to-peer archival storage network that they participate in, through their Prokopius nodes. In a sense, TSSes form a watchdog federation that regulates how the write-once publication medium operates, thereby increasing the accountability of each participant (see Figure 4).

The first basic function of Prokopius is to time stamp grounding values for different TSSes. In that respect, Prokopius acts as a TSS for TSSes, associating a time stamp with the grounding values that different client TSSes submit to it. The functionality of Prokopius is split into two basic categories:

**Publication** A client TSS submits a grounding value for time stamping. When the round is over, Prokopius calculates a time stamp on the client's grounding value and returns it to the client.

**Verification** A client submits a grounding value along with its time stamp for verification. Prokopius looks up the value and returns a yes/no answer along with a proof for that answer.

Although Prokopius exports its time stamping services to all interested TSSes for generic use, it is also its own client so as to provide an identity certificate time stamping and archival service. This service is essential in safeguarding the longevity of signed, time stamped documents beyond the lifetime of the signing key used and of the public key infrastructure within which the key was assigned.

In its capacity as an archive of time stamped identity certificates, Prokopius offers two types of functionality:

**Registration** A client submits a request for archival of its public keys throughout time. In response, Prokopius tracks changes in that client's published public keys, and when it identifies them, it archives previous versions of public keys for that client. The client may, at a later time, request that Prokopius discontinue tracking of its public key.

**Lookup** A client submits a request for the public key of a specific entity at a given time. Prokopius looks up the key and returns it, if found, along with a proof for its existence; if no such key is found, Prokopius returns a proof for the absence of a key for the requested entity at the given time, which means that Prokopius was not tracking that key at the given time.

## 3.2 Example scenarios

In this section we describe a set of simplified examples of the use and operation of Prokopius in the larger setting of services in which it functions. First, we describe the time stamping of a document and its subsequent verification without the help of the original time stamper. Prokopius participates as a high-level TSS and as a secure publication medium.

Then, we explain how the process changes when the document submitted for time stamping is digitally signed, and we want to be able subsequently also to verify the signature, not just the time stamp, on the document. Although Prokopius does not archive digitally signed documents itself, it archives the signing keys that can be used to verify signatures on documents archived through alternate means.

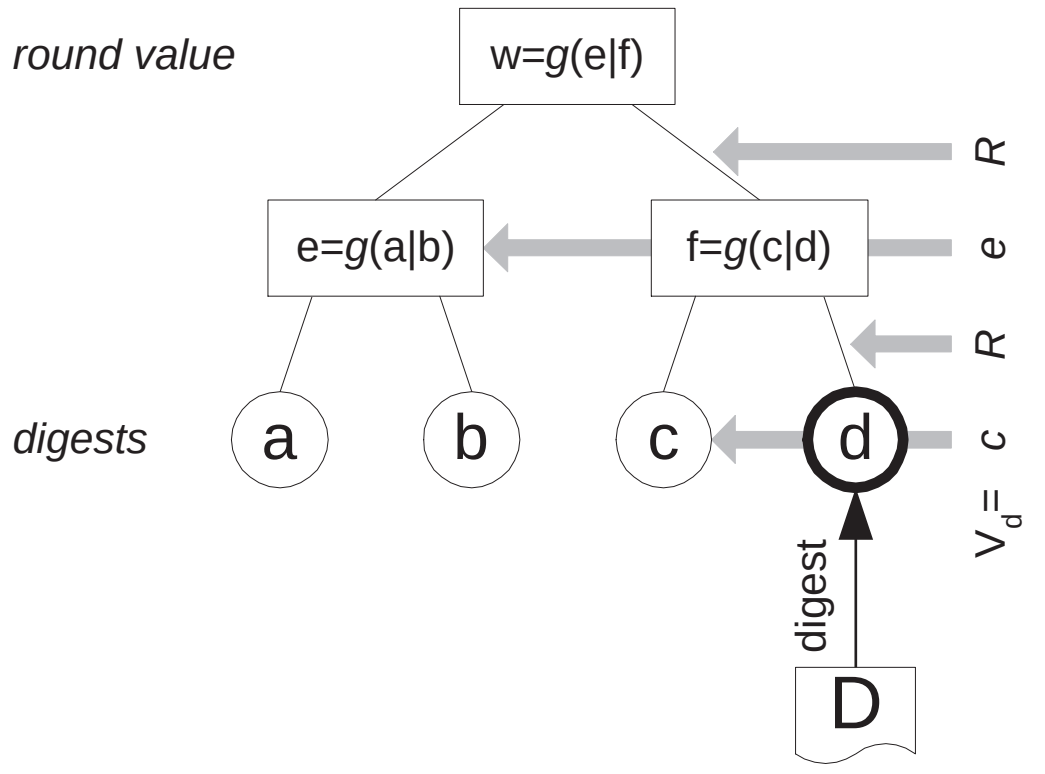

Figure 5: The linking tree built by the TSS $T$ after 13:45:53. Leaves are labeled with the digest of the document to which they correspond. Intermediate nodes are labeled with the result of a collision-resistant, one-way hash function $g(.)$ on the label of the left child node concatenated with the label of the right child node. The concatenation operation is indicated by "|". On the right edge of the figure, the proof of inclusion $V_d$ of document $D$ in the tree is illustrated graphically.

### 3.2.1 Time stamping a document

The author $A$ of document $D$ submits a digest of $D$, $d = digest(D)$ to the Time Stamping Service $T$ for time stamping. Function $digest(.)$ is a well-known, one-way, collision-resistant hash function, which means that it is intractable for anyone to find a different document $D'$, such that $d = digest(D')$. The author could just as well submit the entire document $D$ to the TSS. However, this is not usually done, so as to avoid publication of the contents of $D$, and because $D$ is likely to be much longer than its digest.

When $T$ receives $d$, it holds on to it until the end of its current round. In this example, $d$ arrives at $T$ at 13:45:53 GMT on March 26 1998, and $T$ has second-long rounds. After $T$ has accumulated all document digests that arrived between 13:45:53 and 13:45:54 — digests $a$, $b$ and $c$ in addition to $d$ — it builds a sorted Merkle hash tree out of them, which is the binary linking structure used by $T$ (see Figure 5). Finally, $T$ returns to $A$ the proof $V_d$ of $d$'s inclusion in its linking structure for this time, which consists of the values of $d$'s sibling and the siblings of its ancestor nodes in the tree, as well as a bit at each level indicating whether $d$'s ancestor's position is a left or right child of its parent. The time stamp of $d$ is $V_d = [c/R, e/R]$.

Armed with the time stamped document $\bar{D} = (D, V_d)$, $A$ can prove to $G$ (for *Gullible*, since he trusts $T$) that document $D$ was time stamped at 13:45:54 GMT, on 3/26/1998. All $A$ has to do is give $\bar{D}$ to $G$. $G$, who trusts $T$, first requests the *round grounding value* for 13:45:54 GMT on 3/26/1998 from $T$. In response, $T$ sends back $w$ (refer to Figure 5). Then $G$ knows that if using the proof $V_d$ she can compute the value for $w$, then the time stamp must be valid. She proceeds to calculate $w' = g(e|g(c|digest(D)))$, where "|" declares the concatenation operation. As can be seen from the construction above, $w' = w$, and $G$ is happy with the time stamp.

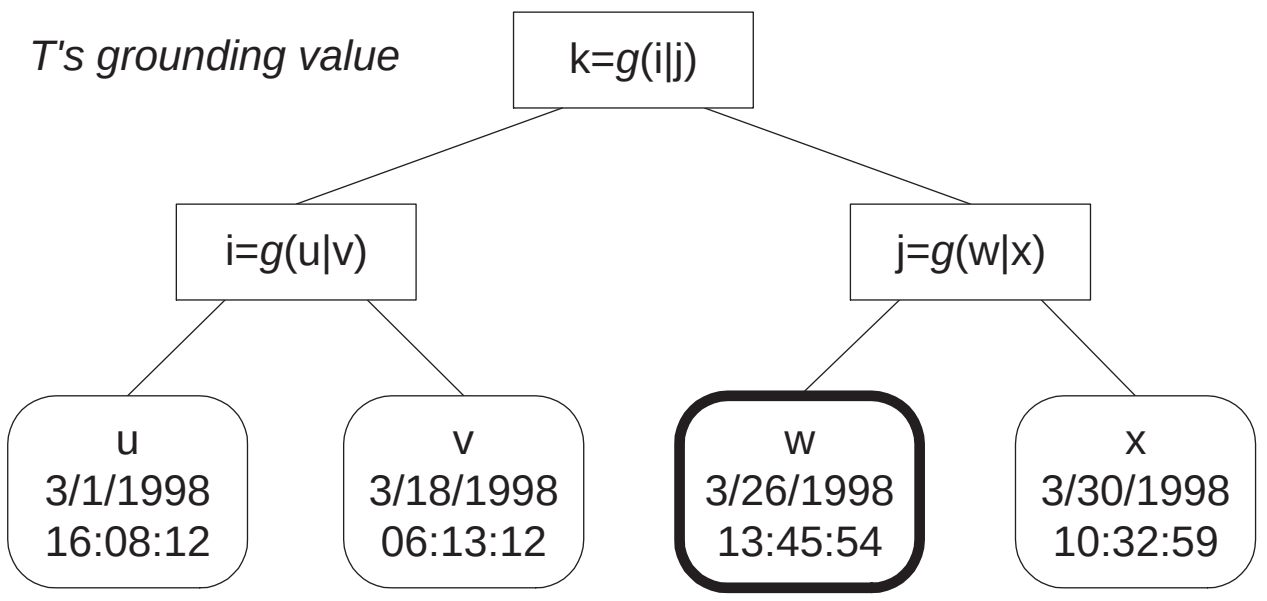

Figure 6: The linking tree built by the TSS $T$ after the end of March 1998, containing all four round values of its four non-empty rounds.

However, since the time stamping service $T$ also wants to be able to convince those who don't trust it, as well as those who do, that the time stamps it issues are correct, it participates in Prokopius. In so doing, $T$ uses Prokopius as a secure, write-once publication medium for a small number of its grounding values. Every month, $T$ puts together all the round grounding values it has accumulated during the month into a linking structure, much like the one used above for digest linking. Figure 6 illustrates the March 1998 linking structure that $T$ builds at the end of the month. March was a fairly slow month, so $T$ had a total of only four rounds during which it received digests for time stamping. $A$ can request from $T$ an extension $V_w$ of the proof of inclusion $V_d$ that it received earlier, showing how document $D$ is linked to the value $k$ that $T$ grounds with Prokopius. As can be seen in Figure 6, $V_w = [x/L, i/R]$.

Now $T$ submits the grounding value $k$ to Prokopius. Prokopius returns to $T$ a proof of inclusion $V_k$ for this grounding value in the global linking structure for March 1998, whose root is the widely published value $l$ ($V_k$ is similar to $V_w$ or $V_d$, so we omit its details; section 5.6 has more information). Through this hierarchical submission to a larger, more widely trusted time stamping entity, $T$ can render its time stamps useful even to verifiers who do not trust it unconditionally. $A$ can request $V_k$ from $T$ and keep it along with the more specific inclu-sion proofs, forming a globally time stamped document $\hat{D} = (D, V_d, V_w, V_k)$. Now $A$ can hand $\hat{D}$ to a verifier $S$ (for *Skeptic*) who trusts Prokopius but not $T$. $S$ first requests a global round grounding value for March 1998 from Prokopius, and receives in response value $l$. Then, $S$ calculates $l'$ from document $D$ and the successive application of proofs $V_d$, $V_w$ and $V_k$ to $D$'s digest. If $l' = l$, then $S$ can accept that $D$ was time stamped some time in March 1998. $\hat{D}$ can be verified regardless of the existence, availability or cooperation of $T$, as long as Prokopius can be reached. If, furthermore, $S$ trusts $T$, he can get the benefit of finer time placement for document $D$.

### 3.2.2 Time stamping a signed document

In this section we expand the time stamping scenario from the previous section to handle signed documents. In this scenario, Prokopius is not only used as the high-level TSS for normal TSSes, but also as an identity certificate archive, used to verify time stamped, signed documents that have been somehow archived.

Using the techniques described in Sections 2.2 and 2.1, archiving a signed document amounts to time stamping the signed document at the time of signing (or at least, before the signing key has changed), and then archiving the associated certificates along with the document. The time stamping portion of the operation is almost identical to that described in Section 3.2.1. The only difference is that instead of submitting the document digest for inclusion in a linking structure at the TSS, $A$ submits his signature $\sigma$ on the document. Since a signature depends on the digest of the document it signs, it is at least as good as the digest itself for building temporal dependence graphs.

More specifically, if $A$ wants to sign document $D$ and archive it for posterity on March 26 1998, he first signs $D$ using his signing key $S_A$, to produce signature $\sigma = s_{S_A}(digest(D))$. Then $\sigma$ is subjected to time stamping at the TSS $T$, to produce $V_\sigma$, as was shown before in Figure 5.

To make the signed document self-sufficient in a long-term archive, $A$ has to include with it his identity certificate $C_A$ also. Recall that $C_A$ is a statement signed by the Certification Authority responsible for $A$, officially associating $A$ with $A$'s public key $S_A$. But then, $C_A$ itself is a signed document that must be archived, and it needs to be time stamped as well. In a practical scheme the certificate $C_A$ would only be time stamped and stored once for all documents signed by $A$ in the period during which $C_A$ is valid. We demonstrate here how this is done for completeness.

As for any other signed document, $C_A$ consists of its data $Z$ (the formal association of the name $A$ with the key $S_A$) and its signature $\sigma' = s_{S_{CA}}(digest(Z))$. $\sigma'$ is now submitted to a time stamping service (for simplicity, we assume it is also submitted to $T$, although it practice it is quite likely that a different TSS is used), yielding the time stamp $V_{\sigma'}$.

Unfortunately, the current master signature verification key $S_{CA}$ for the Certificate Authority also has to be time stamped somehow. This is done with the help of Prokopius. A CA that offers long-term certificate validity to its clients signs up with Prokopius for key change tracking. As part of that service, Prokopius periodically checks the CA's current public key. If the key has changed since the last time it was checked, Prokopius records the time at which this change occurred and the new value of the key. We describe details of how this is implemented in Section 4.

To prepare the signed document for archival, $A$ requests from Prokopius a proof of inclusion $V_{CA}$ of the CA's self-signed key certificate $C_{CA}$ in its archive. This proof is similar to a time stamping proof of inclusion. The time stamped, signed $D$ is now fully represented by $\dot{D} = (D, \sigma, V_\sigma, C_A, V_{\sigma'}, C_{CA}, V_{CA}, V_w, V_k)$. Recall from the previous section that $V_w$ and $V_k$ are the proofs that the TSS $T$ has produced an appropriate globally-grounding value including its current round, and has submitted that to Prokopius successfully.

# 4 Design

In this section we describe the design of Prokopius. We begin with a brief introduction to our two fundamental design goals, longevity and wide trust, that allow Prokopius to operate in settings where current time stamping and archival services cannot. We then break down the basic tasks that Prokopius performs and explain each in detail.

## 4.1 Design goals

The main idea behind Prokopius is to explore the strengths of peer-to-peer distributed systems for building trusted services. To provide the service as outlined in Section 3.2, Prokopius must have the following two properties:

**Longevity** Prokopius must have a high probability of remaining operational for long, continuous time periods (on the order of several decades).

**Wide Trust** Prokopius must be able to convince *at the same time* parties from different areas of the world, with different affiliations, business practices or loyalties — in short, parties who do not trust the same entities — that it performs its tasks correctly.

We achieve longevity by building Prokopius on a loosely-coupled, decentralized distributed system. Such systems exchange complexity for increased reliability and availability. If no system node is more significant than any other node, or equivalently, if the integrity of the system relies equally on all participating nodes, then any and all nodes are replaceable. This replaceability eliminates the threat of single-point short- or long-term failures. Consequently, we can claim informally that, no matter how many participating nodes die, as long as new nodes join at least as frequently as needed, then the system will survive for longer periods than any single, well-maintained node could by itself. We make these descriptions of the conditions under which system longevity can be ensured more specific in the following sections.

To make Prokopius widely trusted, we leverage techniques for achieving distributed consensus in the face of malicious failures (also called *Byzantine* failures). Distributed consensus allows a set of distinct nodes to agree on a particular course of action, based on each node's individual opinion, even when a bounded fraction of all nodes are acting maliciously. In general, this is done by requiring that a large fraction of all nodes (usually, more than two thirds of all nodes) "agree" with a particular operation, before that operation is carried out. Assuming that the number of malicious nodes is relatively small (usually, less than a third of all nodes), this technique allows the overall system to *mask out* the actions of malicious nodes, without even knowing who they are.

Distributed consensus, as well as the related problems of distributed agreement and reliable broadcast, are fairly well-understood but are not in common use in large-scale distributed systems because of their inherent high cost. Prokopius uses distributed consensus techniques only in operations that happen relatively infrequently (at the rate of once a week, or once a month). Therefore, though still expensive, such techniques are affordable in this context and allow Prokopius to meet its wide trust requirement.

## 4.2 Basic tasks

Prokopius operates in rounds of duration $\tau$, following the paradigm of traditional TSSes. Since Prokopius only archives identity certificates, which change infrequently, it need not be designed to react to rapidly changing information, as for example most distributed file systems must do. Furthermore, Prokopius is not a general purpose time stamping service; it only acts as a secure, time stamped publication medium that general purpose time stamping services can use to publish their grounding values. Therefore, $\tau$ can be much larger than the usual duration of time stamping rounds.

To provide the services outlined in Section 3.2, Prokopius must perform the following tasks at every round:

- Track identity certificates for registered identities and update the archive with their time stamped changes (see Section 4.4.2).
- Update its linking structure to include round values that are newly submitted by participating TSSes, as well as its own identity certificate archive round value (see Section 4.4.3).
- Associate a real time with the current round (see Section 4.4.6).

Furthermore, Prokopius must perform the following tasks, for its own self-maintenance:

- Synchronize the change of round among all participating nodes (see Section 4.4.1).
- Update its distributed shared random number generator. This is used in the consensus protocols, and is explained further in later sections (see Section 4.4.4).
- Approve or reject the addition of new nodes in the group, and eject from its ranks nodes that are no longer operational or that no longer enjoy the collective trust of the group (see Section 4.4.5).

Figure 7 shows the high-level break down of each Prokopius round into tasks.

Individually, Prokopius nodes are charged with the responsibility of contributing their independent view of two types of information: first, information that has one *true* value, sometimes called *single source information*, and second, subjective information that is more or less a matter of opinion or local policy, derived separately by each node, and that has no single, true value.

Prokopius nodes must be able to obtain two kinds of single-source information: the current real time, and the latest identity certificates as stored and disseminated by the current Public Key Infrastructures. Though only a single view of each is considered "correct" at any one time, nodes might have inconsistent, outdated views or might maliciously try to misguide others with incorrect views. Using distributed agreement, Prokopius can mask a bounded number of benign or malicious faults in the dissemination of time and identity certificates.

To obtain the current time, each node could use, for instance, an atomic clock, a radio clock or a GPS clock. Note that GPS is actually dependent on the U.S. military, and thus may not be necessarily considered trustworthy by everyone in the world. The nodes do not synchronize their clocks with each other, since this would violate the independence of the clocks of distinct nodes.

There is one kind of subjective information that Prokopius nodes must be able to form: assessments as to which other current or potential fellow Prokopius nodes can be trusted to perform their tasks correctly. These assessments are made based on system observations but also on out-of-band exchanges among Prokopius nodes, and have two goals. First, an assessment must evaluate whether a node in question can meet the system's minimum requirements, such as processing and connection speed, stability, data freshness, etc. Second, an assessment must reflect the belief of the assessor that the target node will dutifully follow the rules of the system. Simplistically, an administrator may form assessments of another node by extrapolating the reputation of the company operating that node. The rule would then be that "good", "solid" companies are likely to run "good", "solid" Prokopius nodes. By forming such assessments, Prokopius nodes effectively form their own watchdog organization. Global assessments, based on the assessments of each participant, are derived using a distributed consensus protocol.

In the following section, we describe in detail the basic assumptions we make for the correct operation of the system.

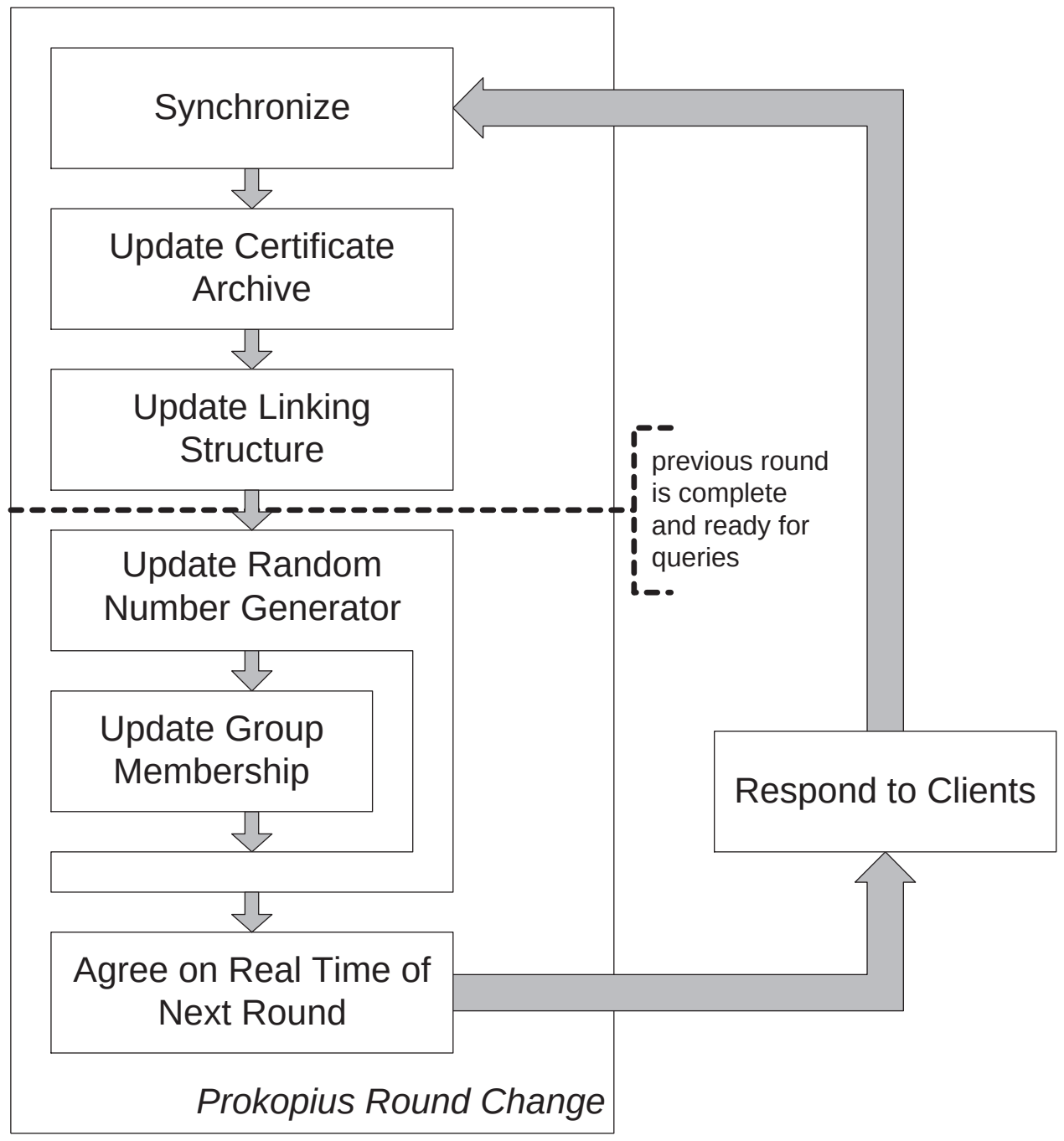

Figure 7: The basic tasks of Prokopius. Round-change tasks are grouped on the left, ordered from top to bottom to indicate their order of execution. The random number generator update task occurs before the membership update task, but completes after it (see Section 4.4.4 for details). Queries are answered throughout the round change process.

## 4.3 System assumptions

The correctness of the design of Prokopius relies on several basic assumptions. Some of these take the form of "minimum system requirements", while others place restrictions on environmental parameters, such as the maximum number of malicious nodes in the system at any one time. We detail all assumptions here, but will also justify each as we go through the specifications of the tasks that necessitate them.

**Assumption 1** *At any one time, no more than $f$ Prokopius nodes are faulty. If the number of all Prokopius nodes at any one time is $n$, then $n \geq 3f + 1$.*

*Faulty* nodes are those nodes that for whatever reason do not follow the Prokopius protocols or those nodes for which the system assumptions in this section do not hold. All other nodes are called *Correct*.

This assumption is drawn from a fundamental theoretical result on distributed agreement [LSP82]. Briefly, there is no protocol that can solve the Byzantine Generals problem (a weaker version of the distributed agreement and consensus problems) if a third or more of all nodes are faulty. It is common to assume this upper bound on faulty nodes, and both the safety and the liveness of the agreement protocols we use depend upon it. However, it is clear that in exceptional cases involving extremely powerful adversaries even such a well-worn assumption can be hard to maintain.

**Assumption 2** *Prokopius nodes fail or lie independently.*

This assumption means that the difficulty with which a group of nodes can be made faulty through means external to the group is proportional to the size of the group. Similarly, a node becomes willingly faulty (i.e., without external coercion) independently of other faulty nodes.

This assumption is important for the distributed consensus and agreement protocols, in conjunction with Assumption 1. A threshold on the number of faulty nodes is only meaningful if the number of faulty nodes reflects the amount of resources needed to cause those nodes to become faulty. Otherwise, more faulty nodes are not necessarily less probable than fewer.

This is the reason why nodes should be in different administrative domains, countries, or other independent organizations. Although complex failure dependences are now being considered by the research community (e.g. [Cac01, MR98]), a satisfactory theory behind inferring arbitrary failure dependencies has yet to be defined.

**Assumption 3** *The current Prokopius nodes, as well as any other entity external to Prokopius, may only perform polynomially bounded, possibly probabilistic computations.*

The implication of this assumption is that current arguably sound cryptographic techniques are safe to use. For example, neither a Prokopius node, nor an external adversary can, with non-negligible probability, find a collision to a collision-resistant hash function.

**Assumption 4** *Prokopius nodes and clients have a well-known mechanism for authenticated communication amongst themselves.*

This mechanism may rely on any particular coexisting infrastructure service (e.g., Verisign, Entrust, DNSSEC), or it may be ad hoc, based on the web-of-trust paradigm or even effected through direct key exchange. This does not mean that the nodes agree on whom they trust. It simply means that correct nodes can retrieve an appropriate, up-to-date public key for any other node in the system.

**Assumption 5** *The effects of a fault on communications do not persist longer than the duration of a Prokopius round.*

For example, we assume that a node will eventually receive messages sent to it in the same round in which they were sent. If an adversary successfully prevents all copies of a message from reaching their destination during an entire round, then we consider the sender of that message to be a faulty node, constrained to the upper limit $f$ imposed in Assumption 1. Large-scale network outages, partitions, and so forth may be caused by the adversary, but they do not persist for the duration of a protocol round. Similarly, no denial of service attack can last longer than a round.

Beyond the constraints described in this assumption, the communication network is completely unpredictable and amenable to coercion by a malicious adversary. Messages may be arbitrarily dropped, reordered, duplicated, etc.

Our timing assumptions are similar in vein to those made by other practical Byzantine fault-tolerant systems[CL00], although we feel they are easier to justify in the less performance-driven context of our application.

**Assumption 6** *All correct nodes have local time sources that lie within $\tau/4$ of the correct global time.*

From this assumption it follows that no two correct nodes' local time sources differ by more than $\tau/2$. This assumption encodes conveniently the reasonable expectation that the local time sources of well-maintained nodes are highly unlikely to drift by more than a few minutes from the correct time, let alone by days. Since $\tau$ is on the order of a couple of days in our setting, imposing a maximum drift of a few hours is not constraining.

Since we never really know exactly what the correct global time is, but merely what our trusted local time source reports, this assumption mandates that correct clocks remain close to the correct global time. This is necessary to support absolute time stamps. For example, if ten years later we ask the system for a certain identity certificate on the 26th of March in 2001, we do not want the query submitter and the system to have very different ideas about what time the 26th of March 2001 actually means. Solutions such as the Timely Computing Base proposal in [VCF00] might help render this assumption practical in all settings and for finer granularities.

**Assumption 7** *No more than $f$ Prokopius nodes are removed from or added into the system in one step.*

This assumption is essential to ensure that Prokopius nodes trust the history of the system. Specifically, by requiring that Prokopius membership change slowly, we require that a newly installed Prokopius node group trusts the group it succeeds. This is because any two successive Prokopius groups share no fewer than $n - f$ nodes, that is no fewer than $n - 2f$ correct nodes, which means that common nodes have certainly more correct than covertly faulty nodes (recall that $n \geq 3f + 1$, and therefore, $n - 2f \geq f + 1$).

## 4.4 Round Change

In this section we present a step-by-step description of the tasks involved in changing rounds (refer to Figure 7 for a high-level view of the operation). The first three steps (Sections 4.4.1, 4.4.2, and 4.4.3) complete the previous round. The last three steps (Sections 4.4.4, 4.4.5, and 4.4.6) prepare and begin the next round.

### 4.4.1 Synchronization

In this step, the nodes come to an agreement that it is time to begin a new round. Each node sets up a *round timer* at the beginning of each round. When its local round timer expires, a node broadcasts a message to all other Prokopius nodes, indicating it is ready to conclude the current round $i$. The message has the form `<Round-Change,`$i$`>`, and bears the signature of the sender. Then, the node waits until $n-2f$ such broadcasts from individual fellow-members, including itself, are received. When a node has received $n - 2f$ it marks the time $\mathcal{S}$ on its local clock, and then proceeds to the remaining steps of the round change.

A node starts listening for `Round-Change` broadcasts before its timer expires, to catch broadcasts from nodes with faster clocks. In fact, immediately after a round change has completed, a node begins listening for the synchronization that signifies the end of the new round. Consequently, a node might actually receive $n - 2f$ synchronization messages for the current round before its own timer has expired. In that case, the node ignores its own timer and proceeds as if its timer had expired, i.e., it marks the local time and moves on to the next step.

Because of assumptions 1 and 6, we know that at least the $n - f$ correct nodes will send out round-change messages, within $\tau/4$ of the correct expiration of the current round. We also know that no more than $f$ of those messages can be dropped or delayed beyond the end of the current round. Therefore, at least $n - 2f$ correct round-change messages will arrive at each Prokopius node, signifying that a majority of all correct nodes are ready to proceed.

### 4.4.2 Certificate archive update

Once a round change has begun, Prokopius nodes go through their local view of the identity certificate archive and update it with recent changes. Two types of changes are recorded in this step. First, nodes agree on any newly arrived registrations or deregistrations of identities. Second, Prokopius checks for new versions of identity certificates for the identities currently being tracked. Any found updates are incorporated into the certificate archive. Both of these subtasks are checked across the entire system for consistency, before being committed. We now describe in detail each of the two subtasks.

During a Prokopius round, clients may send some or all of the Prokopius nodes identity registrations or deregistrations. An identity registration is a signed record of the form `<Register-Identity, [Identity-Name]>`, and signifies the intention of the holder of `[Identity-Name]` to subject his identity certificates to tracking by Prokopius. Similarly, the holder of an identity may send Prokopius nodes a signed deregistration record of the form `<Deregister-Identity, [Identity-Name]>`, signifying that it no longer wishes to have his identity certificates tracked.

Prokopius nodes store the registration and deregistration messages they receive after they have checked the associated signatures and validity. For registrations, the signature is checked using the verification key currently published by the associated public key infrastructure. For deregistrations, the signature is checked using the verification key currently archived by Prokopius for that identity. Finally, registration records for already registered identities, and deregistration records for unregistered identities are discarded.

During the certificate archive update phase of the round change, Prokopius nodes participate in an agreement protocol so as to agree on how the set of tracked identities changes, in response to newly arrived registrations and deregistrations. This is done as follows:

1. Node $j$ puts together a tracking change suggestion record, containing all the valid registrations and deregistrations it has received during the previous round. The suggestion has the form `<Track-Change,` $\mathcal{A}_j$`,` $\mathcal{D}_j$`>`. $\mathcal{A}_j$ and $\mathcal{D}_j$ are the sets of valid registrations and deregistrations respectively that node $j$ has received during the previous round.

2. Node $j$ broadcasts its `Track-Change` record — this record need not be signed by node $j$ — to all Prokopius nodes. Then, it waits until it has collected $n-f$ `Track-Change` records, including its own, from distinct Prokopius nodes. It then removes any registrations or deregistrations whose signature is invalid from every record.

3. Each separate appearance of each registration or deregistration in distinct `Track-Change` records is counted. Node $j$ then puts together a track change proposal $(\bar{\mathcal{A}}_j, \bar{\mathcal{D}}_j)$, a change in the set of tracked identities that was suggested by at least $n - 2f$ Prokopius nodes. This proposal is then submitted to the distributed agreement protocol.

4. The node then waits until agreement is reached, yielding tracking change proposal $(\hat{\mathcal{A}}, \hat{\mathcal{D}})$. The tracking change is applied to the current tracking set, resulting in the set of identities whose identity certificates will be tracked for the round about to close.

The second substage of this task is very similar to the first one. For every identity $\mathcal{I}$ in the set of tracked identities, node $j$ retrieves the corresponding identity certificate $CA_{\mathcal{I}}$ from the associated public key infrastructure. If the retrieved certificate matches the stored certificate, then no further action is taken regarding identity $\mathcal{I}$.

When it has checked all tracked identities, node $j$ puts together an archive update suggestion, similarly to the tracking change suggestion described in step 1 above. The rest of the process proceeds similarly: when node $j$ has accumulated $n-f$ suggestions, it counts the number of distinct suggestions for each certificate update, and then puts together an update proposal, which it submits to another run of the distributed agreement protocol. Once the agreement protocol has terminated, all nodes insert the new certificates into their local archives.

Throughout the discussion of the archive update in this section, we have abstracted away the actual implementation of the certificate archive. Section 5.6 details the particular type of Merkle tree structure we use to maintain snapshots of the archive without excessive storage wastage.

#### 4.4.3 Linking structure update

During this step, Prokopius acts both as a Time Stamping Service, and as a secure, write-once publication medium. As a TSS, Prokopius time stamps the grounding values submitted to it by participating, traditional TSSes, as well as the grounding value of its own identity certificate archive. As a publication medium, it distributes widely the global grounding value of its own linking structure, overseeing its individual nodes to prevent them from distributing information that is inconsistent with its current state.

During the previous round, Prokopius received from participating TSSes individual grounding values. Since Prokopius is designed to operate at time granularities that are much longer than the time granularities of traditional TSSes, it guarantees no temporal precedence among submitted grounding values within the same round. Recall from Section 2.1 that Prokopius can prevent TSSes only from pre/post-dating documents at the larger granularities at which Prokopius itself operates.

Similarly to the process described in the previous step, each node relays the TSS grounding values it has received during the previous round to all other Prokopius nodes. The set of all such valid grounding values received by any Prokopius nodes is then transformed into a linking tree structure, as described in Section 2.1.

The root of the grounding value linking structure is then combined with the root of the archive structure, as it evolved during the previous step, using a one-way collision-resistant function, yielding the *Prokopius round root*. This value represents indisputably (as long as the one-way function remains secure) the archiving and time stamping operations of Prokopius for the finishing round. It is inserted, in turn, into the *Prokopius Time Tree*, yet another linking structure, which is also a search tree keyed by the round number. The new root of the time tree is the current global authenticator $G_i$ for round $i$ of the Prokopius system, and is the principal value used to validate time stamps and archived identity certificates.

Since the contents of the grounding value structure and those of the certificate archive are agreed upon by all nodes, intermediate computations between the derivation of the linking structure and the current global authenticator are performed independently by each node. All intermediate computations are deterministic, so correct nodes should calculate the same global authenticator. Once a Prokopius node has computed the global authenticator for this round, it signs a record containing it of the form `<Global-Authenticator, i, G_i>` and broadcasts it to all other Prokopius nodes. Once a node has received $n - f$ such records, it combines them in a single record bearing the collective signatures of the system (a set of signatures or a threshold signature – see Section 5.2), forming the master global authenticator record `<Master-Global-Authenticator, i, G_i>`, which is then widely distributed, along with the real time association for round $i$ which was determined during the previous round change (see Section 4.4.6 below).

The master global authenticator record is the only value that Prokopius nodes sign collectively for public use, and it is only used for the duration of the following round. This is why Prokopius is immune to the problems of aging and key change that arise with long-term storage of digitally signed documents. The only digital signature that must be trusted by Prokopius clients is derived from current, widely used signing keys, and is only used for periods no longer than perhaps a month.

#### 4.4.4 Shared random number generator update

The purpose of this step is to update the shared random number generator of the system, in anticipation of the impending membership change. This cryptographic construct is used to provide the randomization in the particular Byzantine agreement protocols we use [CKS00] throughout the round change process.

Using the current shared random number generator, Prokopius nodes derive a new random number generator for the upcoming round. The update is necessary because group membership might change from round to round and the particular shared random number generator used is dependent on the current membership. As a result, current Prokopius members prepare the generator change during this step, without yet knowing who is going to remain in the group and who is not. Once the membership has been updated, in the next step, the generator is committed to fit it.

A side effect of the random number generator update process is a record of how different current Prokopius nodes have followed the update protocol. If a node attempts to cheat, hoping to install a predictable shared random number generator, for example, it will be caught with overwhelming probability, and its cheating is provable to others. Such provable "naughty" behavior is used by the nodes in the following step to help weed out faulty nodes from the system.

We describe the particular shared random number generator that we use and its update protocol in Section 5.5.

#### 4.4.5 Membership update

The membership update stage of the round change process is the single, most important step taken by Prokopius to ensure its own longevity. The purpose of this step is to identify participating nodes that no longer function within the fundamental parameters of the system, as outlined in Section 4.3, or no longer function at all. Furthermore, it is the purpose of this step to allow Prokopius to refresh its ranks with new participants that replace those that have been removed.

The overall operation of this step is very similar to the tracking set update stage described in Section 4.4.2. Based on joining requests received during the past round, and on locally derived trust information on current participants, Prokopius nodes make suggestions as to how the system's membership should change.

Every node evaluates properly signed joining requests it has received as well as all current Prokopius members. Each node uses its own independent trust assessment facility described in Section 4.2, to decide whether it has lost trust in some of its fellow nodes, and whether it trusts the potential newcomers to follow the rules as Prokopius members.

Once a complete assessment has been made, the node puts together a membership update suggestion, which it broadcasts to all other nodes, and waits for others' suggestions. Based on the suggestions it receives, it assembles a membership update proposal corroborated by at least $n - 2f$ current Prokopius members, and submits that proposal to the distributed agreement protocol. The membership update proposal agreed upon by all members is then applied to the current membership, and the next round is ready to begin. Note that this method of membership change does not rely on merely *observable* indications of node failure, which can be twisted by an adversary; instead, either *provable* failure evidence is used, or local trust information that is independent of how the protocol runs. This differentiates the Prokopius membership change mechanism from similar techniques that use failure detectors (e.g., [Rei94]) to circumvent the impossibility result of [FLP85].

Honest nodes who receive an agreement that does not include them are expected to go away, and are ignored by others in deliberations associated with the next round. Nodes whose join was just approved request a current copy of the state (the certificate archive and the linking structure) from a current participant. Any current participant will do, since the state is verifiable given the current global grounding value.

One important point about this step is based on As-sumption 7. Since no membership update should change the current membership by more than $f$ nodes, when suggestions are evaluated and proposals are constructed in the process described above, the size of a proposed membership update is always checked against this assumption. This allows the system to maintain its trust continuity, despite membership changes.

To bootstrap the system for the initial round, an initial group membership must be agreed upon through some external means. This might be a political agreement such as allowing each interested government to include a node in the system.

#### 4.4.6 Real time derivation

In this step Prokopius determines a real time association for the next round. In so doing, Prokopius nodes guard the proximity of their communal clock to the global real clock. Assumption 6 ensures that the agreed upon communal time approximates real time within a fraction of the time granularity, specifically within $\tau/4$.

This task proceeds as follows:

1. Every node signs and broadcasts the local time $\mathcal{S}$ at which it synchronized (see Section 4.4.1), in message `<Time-Agree,`$i$`,`$\mathcal{S}$`>`. Given our clock correctness assumptions, no two correct nodes' opinions on the real time are farther apart than $\tau/2$, disregarding network propagation delays.

2. When a node receives $n - f$ *valid* `Time-Agree` messages, it determines the mean $\bar{\mathcal{S}}$ of all $\mathcal{S}$ values. Then the node participates in an instance of the distributed agreement protocol (see Section 5.4), proposing the value $\bar{\mathcal{S}}$ it calculated. The node supports its proposal with a validation that consists of all the `Time-Agree` messages it used to derive its proposal.

3. The node then waits until validated agreement on such a proposal $\hat{\mathcal{S}}$ is reached. Then, $\mathcal{R} = \hat{\mathcal{S}} + \tau$ is the agreed upon real time association for round $i + 1$, and will be widely distributed, along with the global authenticator after the linking update step of the next round change, as described in Section 4.4.5 above.

When a node receives `Time-Agree` messages from other nodes, it has to check their validity. Valid `Time-Agree` messages have correct signatures and are sent from current participants in Prokopius. Before a node stops listening for further `Time-Agree` messages, it checks that among the valid messages it has already received, $n-f$ of them contain $\mathcal{S}$ values that all fall within an interval of width $\tau/2$. If that is not the case, the node continues listening. Assumption 6 ensures that there will eventually be $n - f$ such `Time-Agree` messages.

# 5 Component protocols and data structures

In this section we describe the individual components we use as building blocks for our time stamping service. We list the assumptions made by these protocols and data structures and explain their general purpose.

## 5.1 Authenticated transport layer

We base communications on an authenticated transport layer, such as that provided by Transport Layer Security (TLS), which is a minor modification of the Secure Socket Layer (SSL) [DA99]. We use this protocol to provide authenticated communication between two nodes. As long as a node can verify the public key certificate sent to it by its counterpart (using local certification authority certificates, as per Assumption 4), two nodes can communicate in an authenticated manner. Generally, we use only the Message Authentication Codes (MACs) provided by this protocol, and not signing or encryption. We only use encryption during the update of the shared random number generator (see Section 5.5 for details).

Note that although our goal is to provide permanent authenticated archival of the identity certificates in which the keys used by this component are carried, we do not ourselves provide the associated public key infrastructure service.

## 5.2 Threshold signing facility

In addition to the capabilities afforded by TLS (see Section 5.1), we also use standard threshold signature schemes. Threshold signature schemes are defined within the context of a well-defined group of signers. To be considered *signed*, a message must be individually signed by a number of group members greater than or equal to a preset threshold. This threshold is usually the total number of nodes minus the maximum number of expected misbehaving group members.

We use this facility for all of our broadcast and agreement protocols for simplicity, even though a more efficient implementation might employ a distributed shared signing facility, as seen in the protocol formulations in [CKPS01]. Although distributed shared signing is efficient to use, it generally involves an expensive re-keying procedure when the group membership changes. Ways to perform such renegotiation without resorting to a trusted "key dealer" for different public key cryptosystems have been described [GJKR99, DK01], but we have not yet explored their feasibility for large groups of nodes.

### 5.3 Verifiable consistent broadcast facility

Our agreement protocols require a *verifiable consistent broadcast facility*, in this case, the echo-broadcast facility used in [Rei94] and reformulated for shared signatures in [CKPS01]. The protocol is called *verifiable* because after a sender broadcasts a payload to a group, that sender can compose a *proof* that a sufficient number of group members has actually received the payload. The protocol is also *consistent*, since it ensures that any group members receiving the payload will receive the same payload.

The guarantees of this protocol are weaker than traditional *reliable* broadcast protocols, since it does not ensure that *all* group members receive a broadcast payload; instead, it ensures that, given the maximum expected number of misbehaving group members, enough group members receive a broadcast payload that it is impossible for a different version of the same payload to be accepted by a correct group member. Specifically, if we have $n$ total members and can tolerate no more than $f$ misbehaving members, then each broadcast payload must reach at least $q = \left\lceil \frac{(n+f+1)}{2} \right\rceil$ nodes.

The proof of delivery for a message consists of a set of at least $q$ signatures from group members on a message receipt.

### 5.4 Asynchronous multivalued Byzantine agreement

The objective of this component is to allow a group of participants to come to agreement on a particular decision, even in the face of some malicious participants. It has been theoretically proven that agreement is possible only when the malicious participants are fewer than a third of all participants [LSP82], hence Assumption 1.

We use the validated Byzantine agreement protocol of [CKPS01], which is based on an asynchronous binary Byzantine agreement protocol [CKS00] derived from the seminal randomized agreement algorithm by [BO83]. Randomization is necessary to solve the asynchronous Byzantine agreement problem [FLP85]. Deterministic solutions based on failure detectors have been proposed [Rei94, CL99], however their applicability in our application domain is doubtful; deterministic failure detectors can be subverted by a strong adversary.

Randomization is also essential in addressing the attack of the "aloof malicious participant". In this attack, malicious parties participate in the protocol but over time influence membership changes to include more malicious participants gradually. When the number of malicious participants becomes large enough, Assumption 1, on which all Byzantine agreement protocols rely, becomes false, and the correctness of the protocol fails.

The solution to this problem in practical systems that use deterministic protocols [CL00] is proactive recovery: every once in a while, we go around and "flush" out all nodes, reloading their state and software from a trusted source. This approach indeed helps approximate the assumption that no more than a third of nodes are malicious participants in some situations, but it only works against external intrusion or corruption, and only in a setting where all nodes belong to cooperating administrative organizations. When participating nodes belong to independent organizations, some of which might wish to be malicious, such proactive recovery or software rejuvenation measures do not work.

The randomized algorithm circumvents this problem by using a shared distributed random number generator that is guaranteed to be independent of what the adversary wants, whenever nodes need to break a stalemate while attempting to reach agreement.

### 5.5 Distributed shared random number generator

A *distributed shared random number generator* allows a group of participants to generate communally unpredictable random numbers. In fact, for Prokopius we use a *threshold* random number generator. Such a generator has the property that it needs the participation of at least a number of group members equal to the preset threshold before the shared random number is produced. Using such a generator with a threshold above the maximum potential number of faulty participants $f$ ensures that the random number is not produced before at least one correct participant says it is time to do so. This means that a malicious node has to commit to a particular action in the agreement protocol, without yet knowing what that random number is going to be. It is this feature that makes it hard for a malicious node to influence the outcome of the agreement protocols we use.

Our random number generator (sometimes called a *coin flipping scheme*) draws its security from the as-sumed intractability of calculating discrete logs in finite cyclic groups in polynomial time, and is based on Shamir's secret sharing scheme [Sha79] and an idea first explored in [NPR99] that was further fitted to our communications setting in [CKS00].

To reinitialize the generator when a membership change is impending, we borrow a technique used for the distributed generation of secret signing keys in discrete log based signing systems, originally proposed by Pedersen [Ped91], but most recently amended by Gennaro [GJKR99]. Gennaro's amendment strengthens the original scheme to prevent an adversary from biasing the distribution of the negotiated secret coin signing key. To our knowledge, our work is the first time this scheme has been used to redistribute coin share signing keys in such a coin flipping system.

Briefly, the Pedersen-Gennaro scheme makes every group member create a secret signing key, split it into shares using Shamir's secret sharing [Sha79] scheme, and distribute an encrypted share to each other participant. In the process, the group member produces and publicly broadcasts a set of verification values, that can be used by each secret share recipient to verify the validity of its share. The recipient of a secret share verifies it using the broadcast verification values, keeping track of those secret shares that did not verify accurately.

If a group member fails to verify a secret share, it "accuses" the sender of that share of fraud, and publishes the share it received, along with the corresponding verification values, to prove to the entire group that the sender cheated. Successfully challenged cheaters are excluded from the remaining steps of the protocol. In Prokopius, proven cheaters are suggested for removal from the system by correct nodes, as discussed in Section 4.4.5.

Once all cheaters have been removed, and the consistency of the new group is finalized, every participant combines the shares it received into a secret key, that can be subsequently used to participate in distributed random number generation. In Prokopius, this last step is taken after the membership update stage.

## 5.6 Sorted-tree linking structures

As mentioned in Section 3.2.1, the linking structure we use to represent together a set of documents that are time stamped within the same time period is based on the Merkle tree [Mer80]. A Merkle tree is a regular $k$-ary tree, whose contents are all stored in the leaves, sorted using a predetermined total order. Every internal tree node is labeled by concatenating in order the labels of its $k$ children (or *nil* values for missing children) and applying on the result a one-way, collision-resistant hash function. The label of the root is sometimes called the *root hash* of the tree. The root hash "represents" exactly the ordered set of the leaves of the tree, since to insert or remove one of the leaves of the tree, one would have to find a collision for the hash function. Figure 5 shows a binary Merkle tree, where $g(.)$ is the hash function, $a$, $b$, $c$ and $d$ are the linked documents and $w$ the root hash.

In fact, we use a variation of Merkle trees proposed by Buldas et al. [BLL00], called *authenticated search trees.* The authors suggest this modification to thwart attempts by the maintaining party to keep an inconsistent, unsorted tree linking structure. In these trees data occupy not only leaf nodes, but also internal tree nodes. Furthermore, the computation of a node label takes as input the *search key* of the node in addition to the labels of the node's children. The key property of authenticated search trees is that they allow clients who receive an existence or non-existence proof from the tree maintainer to verify that the maintainer is keeping the tree sorted.

Prokopius uses many different linking trees, for both time stamp linking and certificate archival. We describe how Merkle-like trees accommodate time stamping in Section 3.2.1. In the context of archiving, Merkle-like trees can also be very convenient. Like all trees, they can be efficiently versioned, so as to preserve different snapshots of the set of stored data without excessive redundancy. Figure 8 shows an example of that. The top tree shows the initial version (version 0) of an authenticated search tree. The middle tree shows version 1, resulting from removing the nodes containing $d$ and $k$ from the tree of version 0. The bottom tree is version 2, which results from inserting nodes containing $b$ and $m$ into the tree of version 1. The grayed out nodes are merely references to the original nodes in version 0, and need not be copied for each subsequent snapshot, unless they change in content or label.

Note that in Figure 8, tree operations are *balanced.* This is another welcome property of trees that we use in archiving, since it not only makes for efficient tree update operations, but also keeps existence and non-existence proofs within the linking structure short. Balancing is not necessary in time stamping, since near-complete trees are built from scratch at every round. Balanced trees in Prokopius are red-black trees.

In Prokopius, all tree structures do not have the same properties, among those described above. Figure 9 shows a high-level view of all such trees, at the end of the $n$-th Prokopius round. $A_i$ trees are linking trees put together during round $i$ with all time stamping grounding values submitted by independent TSSes. Nodes are sorted by TSS name and then by time. Such trees are not versioned or balanced, since their nodes are only meaningful within one such $A_i$ tree. $C_i$ trees are certificate archival trees, indicating the version of the certificate archive for round $i$. $C_i$ trees are versioned (so different $C_i$ trees might share common nodes), balanced, and their nodes are sorted by identity name.

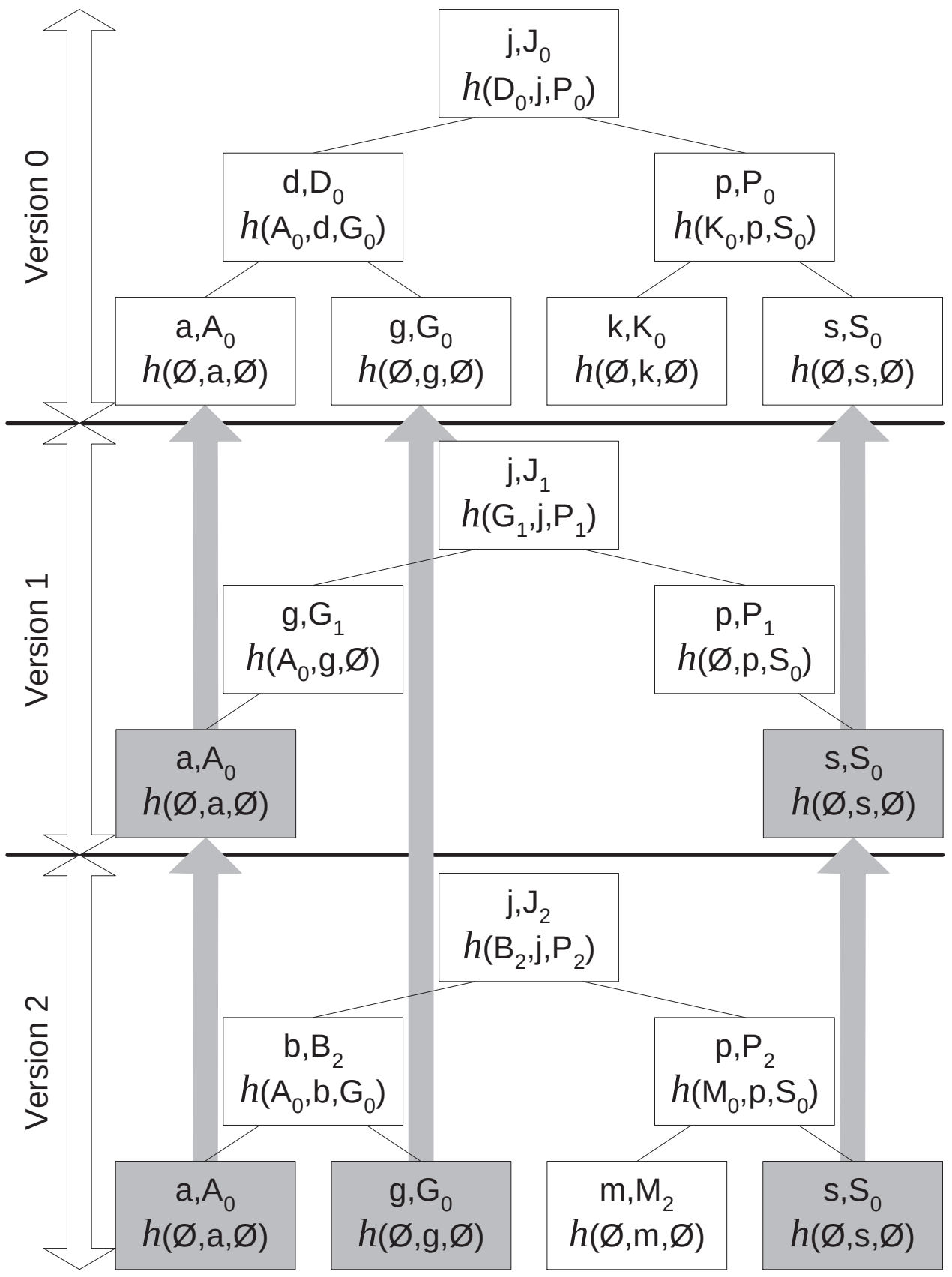

Figure 8: A *versioned*, *balanced* Merkle-Buldas authenticated search tree. Gray nodes are only references to the original nodes to which gray arrows point, and need not be copied in subsequent snapshots of the tree unless they change somehow. Each node is marked with a label/datum pair (e.g., $g, G_0$) and with the hashing operation that produces the label (e.g., $h(\emptyset, g, \emptyset)$).

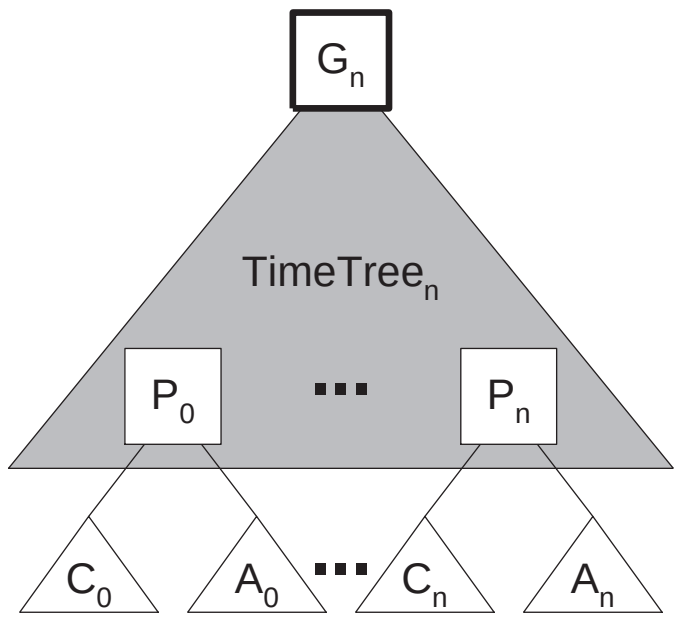

Figure 9: The linking trees of Prokopius during round $n$. $A_i$ trees contain grounding values. $C_i$ trees contain identity certificates. $P_i$ nodes combine the grounding value tree and the certificate tree for round $i$, and participate in the time tree, whose root at round $i$ is $G_i$.

During round $i$, the root hashes of trees $A_i$ and $C_i$ are combined with a one-way collision-resistant hash function and are used to label the *Prokopius root hash* $P_i$ for round $i$. This value represents Prokopius operations during round $i$. $P_i$ nodes are stored in a balanced but not versioned *time tree*. The root hash $G_i$ of the time tree at round $i$ represents present and past operations by Prokopius. $G_i$ is signed by Prokopius nodes and distributed around the world, since it plays the equivalent role of a self-signed root certificate for a Certification Authority. All time stamps and archive identity certificates can be verified through a sequence of existence proofs rooted at $G_i$ during round $i$. In the figure, when Prokopius changes into round $n+1$, a new version of the certificate tree $C_{n+1}$ will be computed as per Section 4.4.2, then a new linking tree $A_{n+1}$ will be computed, as per Section 4.4.3. Trees $C_{n+1}$ and $A_{n+1}$ will be combined into the new Prokopius value $P_{n+1}$, which will be inserted into the time tree, yielding a new root hash $G_{n+1}$, which will be signed by the new Prokopius node membership, and distributed around the world to replace $G_n$. $G_n$ can be discarded and should be considered useless after $G_{n+1}$ has been distributed.

# 6 The Narses simulator

The goal behind the Narses simulator is to simulate a large number of nodes over very long periods of time (on the scale of months) and with highly variable flow sizes. We simulate communication at the flow level, since the computational complexity of packet-level simulators (such as ns [FV99]) is too high for our simulations to run in reasonable time. We wanted a scheduler that could approximate the accuracy of packet-based simulators without the processing time required to consider events on a packet level. To this end, we have attempted to build a discrete event, flow-based simulator that elides individual packet information but still reflects the impact of traffic interdependences. The result is a compromise between speed and accuracy, but one that consistently errs on the side of pessimism (the simulation results report a flow transfer time equal to or greater than what would be consumed in practice.)

Because we make the assumption that all nodes participating in the time stamping service are well-connected, we can reduce the computational complexity of the simulations by assuming that only the links on which the source and destination nodes are attached may limit the bandwidth of a flow. In other words, no intermediate links on a path from the source to the destination other than the first and last hop are bottleneck links. This allows us to model our network topology as a star where every node in the network is connected with a central hub node that has a direct connection to every edge node. Each edge node is assigned a bandwidth that corresponds to an edge host's first-link connection to the Internet. Thus Narses would model a TSS connected to the Internet via a T-1 leased line by assigning that node a bandwidth of 1.544Mbps.

Narses models end-to-end latency by assigning average latencies to each end host. These average latencies can be thought of as the average one way latency from the end host to a central Internet backbone network. When two end hosts communicate, the end-to-end latency is calculated by adding the two hosts' latencies together.

The simulator is modeled after the traditional layered network stack, but we lump the physical, link, network, and transport layers into one layer. The overall behavior of all of these layers acting in unison is modeled using an algorithm we call *fair flow scheduling*, where we define a flow to be all of the bytes of a message passed from a higher layer to the transport layer.

Fair flow scheduling allocates bandwidth to flows only up to their *fair share*, even if the flow could use more than its fair share. The fair share bandwidth is the bandwidth of a node divided by the number of flows sent or received by that node. Where a flow moves between two nodes of different bandwidth, the bandwidth of the flow is determined by taking the *minimum* of the fair share bandwidth of the source and destination of the flow. For example, if a $10 Mbps$ node with four current flows initiates a flow to a $100 Mbps$ node with three current flows, the allocated bandwidth is $\min\{100/(3+1), 10/(4+1)\} = 2 Mbps$. Even though the $100 Mbps$ node could devote $25 Mbps$ to this flow, the flow only takes up $2 Mbps$. In practice this means that some of the other flows on the $100 Mbps$ node could utilize more than 1/4 of the bandwidth, but in our simulation, we pessimistically leave them with their fair share (bandwidth divided by number of flows). As flows are initiated and completed, fair flow scheduling dynamically updates the amount of bandwidth allocated to every active flow in the network. When enough time passes for all of the bits of a flow to be transferred, Narses delivers the flow to the destination node. Essentially, Narses calculates the total transfer time for each flow.

In addition to simulating the transfer times of flows, Narses also simulates a local clock for each node in the simulation. Doing this allows us to model clock drift experienced by real distributed systems. Each node's clock can be configured to drift independently, thus closely approximating a real set of independent clocks. In particular, we use this functionality to verify the parts of Prokopius that must account for clock drift, such as the synchronization protocol described in Section 4.4.1.

The local node clocks in Narses follow a *random linear drift* pattern. Given the maximum offset between the local clock and real time, each node picks a rate of offset change, and a new offset. It then linearly changes its difference from the real time until the new offset is reached. When the new offset is reached, the process is repeated.

Along with a local clock, Narses also offers an interface to simulate operations that take relatively large amounts of CPU time. We use this feature of the simulator to model the delay incurred by performing expensive cryptographic computations, such as generating signatures. Thus Narses offers application and protocol designers a choice between implementing a complete system (which is then faster to port to a deployable system) or approximating complex functionality, which is easier to code and allows the simulations to run faster. Our computation simulations assign to each node a duration of a basic cryptographic operation (a modular multiplication). All other cryptographic operations are broken down roughly into iterations of that basic operation.

All three of the basic node facilities we describe above, network scheduling, clock drift scheduling and heavy computation scheduling, can be flexibly changed to different implementation without affecting the simulated upper layers of the application stack.

## 6.1 Experiment setup

Most network simulators usually have two main aspects of experiment setup: topology and traffic pattern construction. Topology construction creates nodes and connects them with wired links, or in the case of wireless simulations, configures nodes with wireless interfaces. A network stack is also created for each node which simulates the protocols of the network under study. After the topology has been created, traffic is modeled by establishing a number of sources and sinks, and assigning some kind of traffic pattern or distribution to the sources.

Because Narses abstracts away the physical topology of the network, the topology construction phase is lim-ited to defining bandwidth and latency characteristics of end nodes and constructing a stack for each node. At this stage, the experiment also sets up the implementation of the basic node facilities — networking, clock and computation — which in the experiments described in this paper are fair flow scheduling, random linear drift, and modular multiplication-based, as described above.

Apart from the nodes, a Narses experiment defines an *Adversary*. The Adversary is the piece of logic that executes the actual scenario we are testing, in three ways. First, the Adversary may cause nodes to perform legal operations, for example, it may cause the arrival of a new tracking registration at one of the Prokopius nodes. Second, the Adversary may pick arbitrarily a node to subvert; right now, nodes may only be caused to die in our experiments. Third, the Adversary may cause arbitrary network exceptions, such as the loss or delay of a flow, always within the limits imposed by our assumptions in Section 4.3.

When simulator time begins, Narses executes the begin routine of the Adversary, which can set up alarms for further adversarial activity, or perform bootstrapping on the nodes. Once the initialization of the adversary has completed, processing of the event queue begins. When no more node or adversary events occupy the queue, the simulation ends.

Although we built the Narses simulator for evaluating the protocols that comprise our time stamping service, we believe the simulator may be useful for evaluating other large-scale long-running network activities. In separate concurrent work we are evaluating the accuracy and speed of Narses when compared with packet-level simulation. The simulations reported in Section 7 all together took a total time of about 45 minutes running on a dual Pentium 3 Xeon at 1 GHz per processor and one Gigabyte of main memory.

# 7 Evaluation

In this paper we seek to test whether a distributed trusted service using randomized Byzantine agreement can be realistic for our application domain. The slow rate-of-change of information inherent in public key time stamping allows us to use methods that might otherwise be inappropriate.

The basic building block used during every round change is the Asynchronous Multivalued Byzantine Agreement component, which is invoked by members three times to agree on the new group membership for the following round (and by extension, on the exact se-cure shared coin to be used therein), and to agree on the real-time association of a round, (the time that will be included in every time stamp issued during the next round). The agreement has several phases, namely the *echo* phase, where nodes consistently broadcast their validated proposal, the *vote* phase, where nodes broadcast their yes/no vote on a particular proposal by a specific node, and the *agreement* phase, which is the phase during which nodes attempt to reach consensus on whether or not to use a particular validated proposal or not.

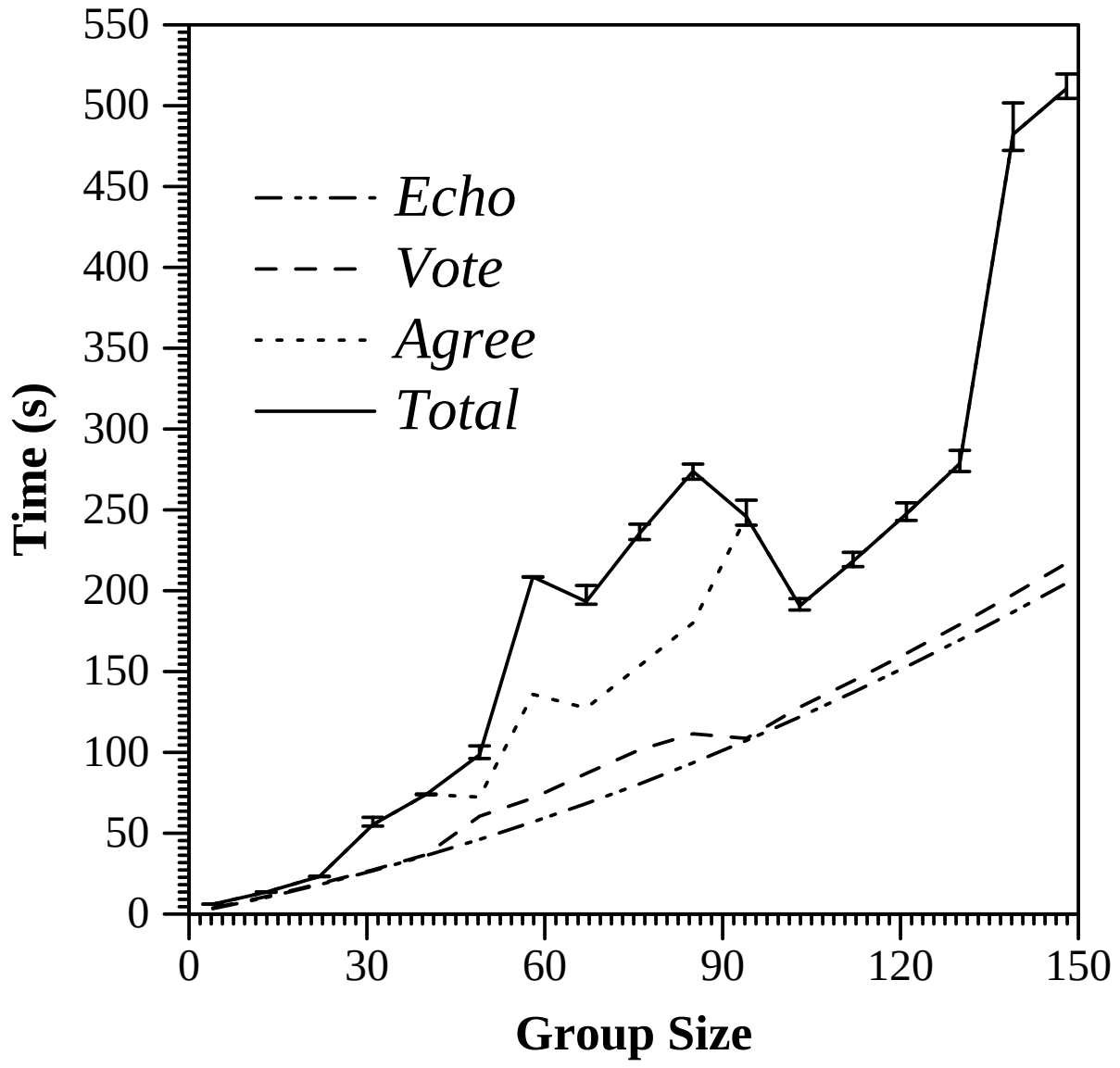

Figure 10: This graph shows the increase in the completion time of the different phases of a single multivalued Byzantine agreement protocol instance. The lowest curve shows the proposal broadcast (echo) phase, The second lowest curve shows the additional time for the vote broadcast phase. The third curve shows the additional time to complete a single binary agreement. The top curve shows the total time taken by the multivalued agreement.

Figure 10 presents the results of our simulations. This graph shows the increase in the completion time of the different phases of a single Multivalued Byzantine Agreement protocol instance as the group membership size increases. The maximum group membership size simulated is 148 nodes. The lowest curve shows the echo phase, which happens exactly once per agreement. The second lowest curve shows the additional time taken for the vote broadcast phase, which is much shorter than the proposal (echo) broadcast phase, because it does not require a consistent broadcast. The third curve shows the additional time taken to complete a single binary agreement. The top curve shows the total time taken by the multivalued agreement. The discrepancy between the total time and the time to complete the binary agreement is due to the fact that the vote and binary agreement phases may happen more than once, depending on the proposal chosen to vote on. Since the adversary knows in advance in which order the proposals are voted on, it can schedule the corruption of the first $f$ proposers, causing the eventual agreement to delay due to repetition of the voting and binary agreement phases.

However, the adversary cannot cause more than $2f$ such repetitions [CKPS01]. Given 148 nodes, this means $2 \times 49$ total possible repetitions of the voting and agreement phases for a single multivalue agreement. From the graph, we see that these phases together account for about 510 seconds for a single repetition. Adding the 204 seconds required for the echo phase, a single multivalue agreement may take about 14 hours. Since three such multivalue agreements are required for one round, this means the adversary can cause a time stamping round to take about 42 hours, or almost two days.

This is indeed a worst-case scenario, for two reasons. First, although we have not done so, the membership agreement and real-time association agreement steps of the time stamping protocol may in fact be combined, requiring only two multivalue agreements per round. Second, Cachin reports a protocol with a constant number of expected rounds [CKPS01] for multivalued agreement which also relies on randomization. However, we have not yet implemented or simulated this protocol.

Thus it is reasonable to state that a single round of the time stamping protocol for 148 nodes with an adversary causing maximal damage can still be performed in less than a couple of days.

# 8 Related work

Although time stamping seems to be essential to conduct secure transactions with lasting effects in the digital world of the Internet, the number of researchers exploring this topic is surprisingly small. In addition to the time stamping work described in Section 2.1, most notably Just [Jus98] describes the intricacies of relying on the trustworthiness of the time stamper to ensure the immutability of the document time line, and ways in which earlier work fails to make time stampers accountable. This work, although not directly applicable to the core time stamping described in this paper (where relative order of public keys within a round is insignificant), is very important for the operation of the potential document time stamping services run by the "edges" of our network, by the group member nodes individually. A very detailed specification of a time stamping service has been produced in the TIMESEC project [QMA+99].

Practical systems that tolerate Byzantine faults have been proposed and implemented in the past (e.g., [CL99, Rei94]). However, only deterministic approaches have been used, due to the fact that randomization was traditionally considered expensive to use. To circumvent the impossibility result of [FLP85], various avenues are followed, including the use of failure detectors that are very hard to build and prove correct in the case of systems where group membership must change, or the imposition of a static membership that is preconfigured by the system's administrators. Unfortunately, we do not have the luxury of knowing well in advance what the projected membership of our system's group will be far into the future, nor can we assume that a single administrative organization can be charged with maintaining this scheme; in fact, we most certainly do not want a single organization to perform this task all by itself. It seems that in such a setting, where we can use neither proactive recovery [CL00], nor a static group membership, randomization is the only way to go. Fortunately, the timing requirements of our application are lax enough that we can employ the more expensive randomized agreement techniques described here without significant service quality reduction. Recently, Cachin [Cac01] described an architecture for secure and fault-tolerant service replication on the Internet using the basic components we use in this work, although the work relies on the group membership being static.

Other researchers have explored techniques for extracting out detail and hence reducing computation complexity for simulations of large-scale networks. Flowsim [ADET93] abstracts away some of the detail of packet-level simulators by grouping packets with closely spaced send times together into a single event called a packet train. This is motivated by the observation that links on real networks are frequently occupied solely by packets from the same flow.

In addition, [HEH98] develops two abstractions for simulating large scale multicast networks. The first abstraction is to eliminate maintenance messages that maintain the multicast tree. The second abstraction, and the one most related to our work, is that they do away with hop-by-hop routing and instead schedule packet delivery directly from source to destination. The key difference is that they do not take queuing delay into account, which means that cross traffic does not affect the transmission of packets. We take into account queuing delay by dividing up bandwidth between concurrent flows in or out of the same node. However, we also ignore cross traffic that is unrelated to Prokopius, and pretend there is no other service running in the world.

# 9 Future work

There are several avenues of future work for us. Some obvious areas are proof of correctness of the protocols, further validation of the Narses simulator, and evaluation of a reasonably-sized deployment of Prokopius.

While Prokopius is merely a TSS for public key information, as described in Section 2.1 it is trivial to extend it to an arbitrary document time stamping service. In this sense, Prokopius becomes the secure publication medium for locally provided time stamping services that can lie at the edge of the network. We plan to extend Prokopius in this way, since we require a document TSS for the following project.

Prokopius is but one component of a larger project focused on providing *histories of online names*. An online name might be an email address, instant messaging identifier, or perhaps even a telephone number. It is likely to be an identifier that works over several applications or online media. Such names are useful for several purposes, including contacting people or assigning identities to online contracts. While there are several projects and companies that aim to provide unique online identifiers for people ([RKJ00, MB00] or `www.onename.com`), we recognize that people's online names are likely to change due to changes in employment or geographical address or because the companies hosting the online names go out of business or themselves change names. Knowing an old online name for a person is thus no help in reaching that person if there is no way to find the new name from the old one.

Assuming anyone wishes to do so, our naming history service allows people to provide an authenticated history of their online names, which is a list of signed mappings between each name and the next most recent name. When given an old name (and approximate time frame during which that name was valid), the history service returns the newest online identifier for that person. The history service is an example of authenticated archived information. Mappings between an old name and a newer name in the service may be authenticated using a public key that is no longer valid. The time stamping service is thus an essential component that allows us to validate the authentication of each link in a person's naming history, and we would like to finish building the history service and evaluate Prokopius in that context.

# 10 Conclusion

In this paper we show that it is possible to build secure reliable and survivable services in a peer-to-peer network using Byzantine fault-tolerant protocols. Prokopius is able to offer public key time stamping rounds on the order of a couple of days for a 148-node network even in the case where an adversary causes the maximum damage allowed within our fault model. Rounds on the order of a couple of days, or even on the order of a week, are sufficient to support applications, such as archival of public key snapshots, where the information itself changes slowly.

In contrast to a centralized time stamping service, the distributed service can survive changes in public keys and even a complete change in service provider membership over time. The service can validate documents signed in the past with keys no longer in service and by entities that have ceased to exist.

This public key time stamping service can also form the core of a time stamping service for other kinds of documents, also built in a peer-to-peer network. In this case, individual nodes in the network may time stamp the documents submitted to them without consultation with other nodes, and thus with only local overhead for each time stamp. Every once in a while, these nodes submit the round hashes of their individual time stamping efforts to the group for agreement.

## 11 Acknowledgements

This work is supported by the Stanford Networking Research Center, and by DARPA (contract N66001-00-C-8015). Petros Maniatis is supported by a USENIX Scholar Fellowship.

The research included in this work has benefitted greatly from discussions within the peer-to-peer networking research group at Stanford, as well as with Kevin Lai, Dan Boneh, Patrick Lincoln and Mema Roussopoulos. We thank them for their invaluable help.

## References

[ADET93] Jong-Suk Ahn, Peter B. Danzig, Deborah Estrin, and Brenda Timmerman. Hybrid technique for simulating high bandwidth-delay computer networks. In *Proceedings of the 1993 ACM SIGMETRICS conference on Measurement and modeling of computer systems*, pages 260–261, Santa Clara, CA, USA, May 1993. ACM SIGMETRICS, ACM Press.

[BdM91] Josh Benaloh and Michael de Mare. Efficient broadcast time-stamping. Technical Report TR-MCS-91-1, Clarkson University, Department of Mathematics and Computer Science, April 1991.

[BLL00] Ahto Buldas, Peeter Laud, and Helger Lipmaa. Accountable certificate management using undeniable attestations. In *Proceedings of the 7th ACM Conference on Computer and Communications Security*, pages 9–17, Athens, Greece, November 2000. Association for Computing Machinery, ACM Press.

[BLLV98] Ahto Buldas, Peeter Laud, Helger Lipmaa, and Jan Villemson. Time-stamping with Binary Linking Schemes. In Hugo Krawczyk, editor, *Advances on Cryptology — CRYPTO '98*, volume 1462 of *Lecture Notes in Computer Science*, pages 486–501, Santa Barbara, USA, August 1998. Springer Verlag.

[BO83] Michael Ben-Or. Another advantage of free choice: Completely asynchronous agreement protocols. In *Proceedings of the 2nd ACM Symposium on Principles of Distributed Computing (PODC)*, pages 27–30, Montreal, Canada, August 1983. ACM Press.

[Cac01] Christian Cachin. Distributing trust on the Internet. In *Proceedings of the International Conference on Dependable Systems and Networks (DSN-2001)*, Göteborg, Sweden, June 2001. IEEE.

[CKPS01] Christian Cachin, Klaus Kursawe, Frank Petzold, and Victor Shoup. Secure and efficient asynchronous broadcast protocols. Technical Report RZ 3317, IBM Research, Zurich, Switzerland, March 2001.

[CKS00] Christian Cachin, Klaus Kursawe, and Victor Shoup. Random oracles in Constantinople: Practical asynchronous Byzantine agreement using cryptography. In *Proceedings of the 19th ACM Symposium on Principles of Distributed Computing (PODC 2000)*, pages 123–132, Portland, OR, USA, July 2000. ACM.

[CL99] Miguel Castro and Barbara Liskov. Practical Byzantine fault tolerance. In *Proceedings of the 3rd Symposium on Operating Systems Design and Implementation (OSDI 1999)*, New Orleans, LA, USA, February 1999. USENIX Association.

[CL00] Miguel Castro and Barbara Liskov. Proactive recovery in a Byzantine-fault-tolerant system. In *Proceedings of the 4th Symposium on Operating Systems Design and Implementation (OSDI 2000)*, pages 273–287, San Diego, CA, USA, October 2000. USENIX Association.

[CSWH00] Ian Clarke, Oskar Sandberg, Brandon Wiley, and Theodore W. Hong. Freenet: A distributed anonymous information storage and retrieval system. In Hannes Federrath, editor, *Proceedings of the Workshop on Design Issues in Anonymity and Unobservability*, volume 2009 of *Lecture Notes in Computer Science*, pages 46–66, Berkeley, CA, USA, July 2000. International Computer Science Institute (ICSI), Springer Verlag.

[DA99] T. Dierks and C. Allen. RFC 2246: The TLS protocol version 1, January 1999. Status: PROPOSED STANDARD.

[DFM00] Roger Dingledine, Michael J. Freedman, and David Molnar. The Free Haven project: Distributed anonymous storage service. In Hannes Federrath, editor, *Proceedings of the Workshop on Design Issues in Anonymity and Unobservability*, volume 2009 of *Lecture Notes in Computer Science*, pages 67–95, Berkeley, CA, USA, July 2000. International Computer Science Institute (ICSI), Springer Verlag.

[DK01] Ivan Damgård and Maciej Koprowski. Practical threshold RSA signatures without a trusted dealer. In *Proceedings of the International Conference on the Theory and Application of Cryptographic Techniques (EUROCRYPT 2001)*, volume 2045 of *Lecture Notes in Computer Science*, pages 152–165, Innsbruck, Tyrol, Austria, May

2001. International Association for Cryptologic Research, Springer Verlag.

[FLP85] Michael Fischer, Nancy Lynch, and M. Paterson. Impossibility of distributed consensus with one faulty process. *Journal of the Association for Computing Machinery*, 32(2):374–382, April 1985.

[FV99] K. Fall and K. Varadhan. *ns notes and documentation*. The VINT Project, July 1999. available at http://www-mash.cs.berkeley.edu/ns/.

[GJKR99] Rosario Gennaro, Stanisław Jarecki, Hugo Krawczyk, and Tal Rabin. Secure distributed key generation for discrete-log based cryptosystems. In Jacques Stern, editor, *Proceedings of the International Conference on the Theory and Application of Cryptographic Techniques (EUROCRYPT 99)*, volume 1592 of *Lecture Notes in Computer Science*, pages 295–310, Prague, Czech Republic, May 1999. International Association for Cryptologic Research, Springer Verlag.

[Gnu] Gnutella. http://www.gnutellanews.com/.

[HEH98] Polly Huang, Deborah Estrin, and John Heidemann. Enabling large-scale simulations: Selective abstraction approach to the study of multicast protocols. In *Proceedings of the Sixth International Symposium on Modeling Analysis and Simulation of Computer and Telecommunications Systems (MASCOTS '98)*, pages 241–248, Montreal, Canada, July 1998. IEEE Computer Society, IEEE Press.

[HS91] Stuart Haber and W. Scott Stornetta. How to time-stamp a digital document. *Journal of Cryptology: the journal of the International Association for Cryptologic Research*, 3(2):99–111, 1991.

[Jus98] Michael Just. Some timestamping protocol failures. In *Proceedings of the Symposium on Network and Distributed Security (NDSS 98)*, San Diego, CA, USA, March 1998. Internet Society.

[KBC+00] John Kubiatowicz, David Bindel, Yan Chen, Steven Czerwinski, Patrick Eaton, Dennis Geels, Ramakrishan Gummadi, Sean Rhea, Hakim Weatherspoon, Westley Weimer, Chris Wells, and Ben Zhao. Oceanstore: an architecture for global-scale persistent storage. In *Proceedings of the 9th international conference on Architectural support for programming languages and operating systems (ASPLOS 2000)*, pages 190–201, Cambridge, MA, USA, November 2000. ACM, ACM Press.

[LSP82] Leslie Lamport, Robert Shostak, and Marshall Pease. The Byzantine generals problem. *ACM Transactions on Programming Languages and Systems*, 4(3):382–401, July 1982.

[LV99] Arjen K. Lenstra and Eric R. Verheul. Selecting cryptographic key sizes. http://www.cryptosavvy.com/cryptosizes.pdf, November 1999.

[MB00] Petros Maniatis and Mary Baker. Identiscape: Tackling the personal online identity crisis. Technical Report CSL-TR-00-804, Computer Systems Laboratory, Stanford University, Stanford, CA, USA, June 2000.

[Mer80] Ralph C. Merkle. Protocols for public key cryptosystems. In *Proceedings of the 1980 Symposium on Security and Privacy*, pages 122–133, Oakland, CA, U.S.A., April 1980. IEEE Computer Society.

[MR98] Dahlia Malkhi and Michael Reiter. Byzantine quorum systems. *Distributed Computing*, 11(4):203–213, 1998.

[NPR99] Moni Naor, Benny Pinkas, and Omer Reingold. Distributed pseudo-random functions and KDCs. In Jacques Stern, editor, *Proceedings of the International Conference on the Theory and Application of Cryptographic Techniques (EUROCRYPT 99)*, volume 1592 of *Lecture Notes in Computer Science*, pages 327–346, Prague, Czech Republic, May 1999. International Association for Cryptologic Research, Springer Verlag.

[Ped91] Torben P. Pedersen. A threshold cryptosystem without a trusted party. In *Proceedings of the International Conference on the Theory and Application of Cryptographic Techniques (EUROCRYPT 1991)*, volume 547 of *Lecture Notes in Computer Science*, pages 522–526, Brighton, UK, April 1991. International Association for Cryptologic Research, Springer Verlag. Extended Abstract.

[QMA+99] J. J. Quisquater, Henri Massias, J. S. Avilla, B. Preneel, and B. Van Rompay. TIMESEC: Specification and implementation of a timestamping system. Technical Report WP2, Université Catholique de Louvain, 1999.

[Rei94] Michael Reiter. Secure agreement protocols: Reliable and atomic group multicast in Rampart. In *Proceedings of the 2nd ACM Conference on Computer and Communications Security*, pages 68–80, Fairfax, VA, USA, November 1994. Association for Computing Machinery, ACM Press.

[RKJ00] Bhaskaran Raman, Randy H. Katz, and Anthony D. Joseph. Universal inbox: Providing extensible personal mobility and service mobility in an integrated communication network. In *Proceedings of the Workshop on Mobile Computing Systems and Applications (WMSCA 2000)*, pages 95–106, Monterey, CA, USA, December 2000. USENIX Association and IEEE.

[Sha79] Adi Shamir. How to share a secret. *Communications of the ACM*, 22(11):612–613, November 1979.

[Sur] Surety, Inc. Secure time/date stamping in a public key infrastructure. available at http://www.surety.com/.

[VCF00] Paulo Veríssimo, António Casimiro, and Christoff Fetzer. The timely computing base: Timely actions in the presence of uncertain timeliness. In *Proceedings of the International Conference on Dependable Systems and Networks (DSN 2000)*, pages 553–542, New York, NY, USA, June 2000. IEEE Computer Society, IEEE Press.

[WRC00] Marc Waldman, Aviel Rubin, and Lorrie Cranor. Publius: A robust, tamper-evident, censorship-resistant web publishing system. In *Proceedings of the 9th USENIX Security Symposium*, pages 59–72, Denver, CO, USA, August 2000. USENIX Association, USENIX Association Press.

`$Id: TechReport.tex,v 1.50 2001/06/28 22:56:31 maniatis Exp $`